\documentclass[%
aip,
amsmath,amssymb,
preprint,%
author-year,%
]{revtex4-2}

\usepackage{graphicx}%
\usepackage{rotating}
\usepackage{dcolumn}%
\usepackage{bm}%
\usepackage[mathlines]{lineno}%

\usepackage[utf8]{inputenc}
\usepackage[T1]{fontenc}
\usepackage{mathptmx}
\usepackage{etoolbox}

\usepackage{epsfig}

\usepackage{nomencl}%

\usepackage{amssymb}
\usepackage{amsthm}

\usepackage{lineno}

\usepackage{etex}
\usepackage{graphicx}
\usepackage{epstopdf, epsfig}
\usepackage[T1]{fontenc}
\usepackage{tikz}
\usepackage{calc}

\usepackage{amssymb}
\usepackage{booktabs}

\usepackage[english]{babel}
\usepackage[T1]{fontenc}
\usepackage[utf8]{inputenc}
\usepackage{tabu}
\usepackage[usestackEOL]{stackengine}
\setstackgap{L}{\normalbaselineskip}
\usepackage{soul} %
\usepackage{notoccite} %
\usepackage{mathtools} %

\usepackage{svg}
\usepackage[hidelinks]{hyperref}

\usepackage{cleveref}
\crefname{figure}{Fig.}{Fig.}
\crefname{equation}{Eq.}{Eq.}

\usepackage{bm}

\usepackage{placeins}

\usepackage{cancel}

\usepackage{clipboard}%

\usepackage{enumitem}

\newcommand{\ie}{i.e.\ }%

\newcommand{\re}{Re}
\newcommand{\str}{St}

\newcommand{\lam}{\lambda}

\newcommand{\uinf}{u'_{\rm \infty}}

\newcommand{\ctm}{\bar{C}_{\rm T,pair}}
\newcommand{\ctms}{\bar{C}_{\rm T,single}}

\newcommand*\diff{\mathop{}\!\mathrm{d}}

\newcommand{\etamp}{\eta_{\rm pair}/\eta_{\rm single}}

\newcommand{\caseNum}{550}

\newcommand{\etaamp}{\eta_{\rm pair} / \eta_{\rm single}}

\makeatletter
\def\@email#1#2{%
	\endgroup
	\patchcmd{\titleblock@produce}
	{\frontmatter@RRAPformat}
	{\frontmatter@RRAPformat{\produce@RRAP{*#1\href{mailto:#2}{#2}}}\frontmatter@RRAPformat}
	{}{}
}%
\makeatother
\begin{document}
	
	\preprint{XXXX-XXXX}
	
	\title[]{In-phase schooling hinders linear acceleration in wavy hydrofoils paired in parallel across various wavelengths}

	\author{Zhonglu Lin}
	\affiliation{Key Laboratory of Underwater Acoustic Communication and Marine Information Technology of the Ministry of Education, College of Ocean and Earth Sciences, Xiamen University, Xiamen City, Fujian Province, 361005, China}%
	\affiliation{Engineering Department, University of Cambridge, Cambridge City, Cambridgeshire, CB2 1PZ, United Kingdom}%
	\affiliation{State Key Laboratory of Marine Environmental Science, College of Ocean and Earth Sciences, Xiamen University, Xiamen City, Fujian Province, 361005, China}
	
	\author{Dongfang Liang}
	\affiliation{Engineering Department, University of Cambridge, Cambridge City, Cambridgeshire, CB2 1PZ, United Kingdom}%
	\email{dl359@eng.cam.ac.uk}

	\author{Yu Zhang}%
	\affiliation{Key Laboratory of Underwater Acoustic Communication and Marine Information Technology of the Ministry of Education, College of Ocean and Earth Sciences, Xiamen University, Xiamen City, Fujian Province, 361005, China}%
	\affiliation{State Key Laboratory of Marine Environmental Science, College of Ocean and Earth Sciences, Xiamen University, Xiamen City, Fujian Province, 361005, China}
	\email{yuzhang@xmu.edu.cn}

	\date{\today}%

	\begin{abstract}
		This study examines the impact of in-phase schooling on the hydrodynamic efficiency during linear acceleration in a simplified model using two undulating NACA0012 hydrofoils arranged in phalanx formation as a minimal representation of a fish school. The research focuses on key parameters, namely Strouhal (0.2-0.7) and Reynolds (1000-2000) numbers, and explores the effect of varying fish-body wavelengths (0.5-2), reflecting natural changes during actual fish linear acceleration. Contrary to expectations, in-phase schooling did not enhance acceleration performance. Both propulsive efficiency and net thrust were found to be lower compared to solitary swimming and anti-phase schooling conditions. The study also identifies and categorizes five distinct flow structure patterns within the parameters investigated, providing insight into the fluid dynamics of schooling fish during acceleration.
	\end{abstract}

	\maketitle
	
	\section{Introduction}
	
	Fish species are commonly observed in aquatic environment. Its swimming dynamics has been extensively researched across disciplines like anatomy, animal behavior, robotics, and computational simulation \citep{Webb1984, weihs1973hydromechanics, Chao2021, Ashraf2017, Li2020, Borazjani2010, Chao2019}. These studies aim to understand fish locomotion and inspire the development of autonomous underwater vehicles (AUVs) with superior performance. However, current AUVs fall short compared to natural swimmers in many aspects, including manoeuvrability, speed, efficiency and stealth \citep{Fish2020}.
	
	This study investigates how the wavelength affects the swimming behaviour of two side-by-side undulating NACA0012 hydrofoils. Understanding these effects can shed light on the mechanisms behind accelerated fish schools. Acceleration in fish schools serves purposes like predator evasion and collective maneuvers, utilizing sensory inputs such as vision, lateral line sensing, and proprioception \citep{Partridge1981, Zheng2005, Deng2021, Lecheval2018, Rosenthal2015, Coombs2014, Li2021a}. Although the effects of wavelength on individual swimmers have been studied \citep{Thekkethil2017, Khalid2021}, its influence on accelerating fish schools remains largely unexplored \citep{lin2022swimming,lin2022wavelength}.
	
	This study addresses the research gap by reviewing existing literature on fish swimming, specifically focusing on the effects of wavelength, acceleration, and side-by-side fish schooling. Some previous studies have used simplified cylinder representations, which fail to capture the full fluid-structure interaction of fish schooling dynamics \citep{lin2022flow, Lin2019, Lin2018c, Lin2018b, Lin2016a, Lin2017a, Gazzola2012, NAIR2007, Hlamb1932hydrodynamics}.

	In nature, the wavelength of swimming bodies varies across species and locomotion phases \citep{Santo2021, DuClos2019}, as seen in \Cref{fig:4modes}. \cite{Santo2021} studied 44 body-caudal fin (BCF) fish species and found that the median wavelength significantly increased from anguilliform to thunniform swimming styles. The wavelength range for the tested species was 0.5 to 1.5 times the body length. The overlap in wavelengths suggests compatibility with different body shapes and swimming styles. \cite{DuClos2019} observed that the wavelength during escape acceleration was approximately twice as long as during steady swimming, being 2 times of the body length. Accordingly, the present study chose the wavelength to vary from 0.5 to 2.0 body lengths.

	\newcommand{\addlabeltrim}[3]{%
		\begin{tikzpicture}
			\node[anchor=south west,inner sep=0] (image) at (0,0) 
			{\includegraphics[width=#1\linewidth, trim={0cm 22cm 0cm 22cm},clip]{{{#2}}}};%
			\begin{scope}[x={(image.south east)},y={(image.north west)}]
				\node[anchor=south west] at (0.00,0.75) {\footnotesize #3};	%
			\end{scope}
		\end{tikzpicture}%
	}
	\newcommand{\addlabelintro}[3]{%
		\begin{tikzpicture}
			\node[anchor=south west,inner sep=0] (image) at (0,0) 
			{\includegraphics[width=#1\linewidth]{{{#2}}}};%
			\begin{scope}[x={(image.south east)},y={(image.north west)}]
				\node[anchor=south west] at (0.3,0.9) {\footnotesize #3};	%
			\end{scope}
		\end{tikzpicture}%
	}
	\newcounter{testdd}
	\setcounter{testdd}{0}
	\newcommand\counterdd{\stepcounter{testdd}\alph{testdd}}
	\begin{figure*}
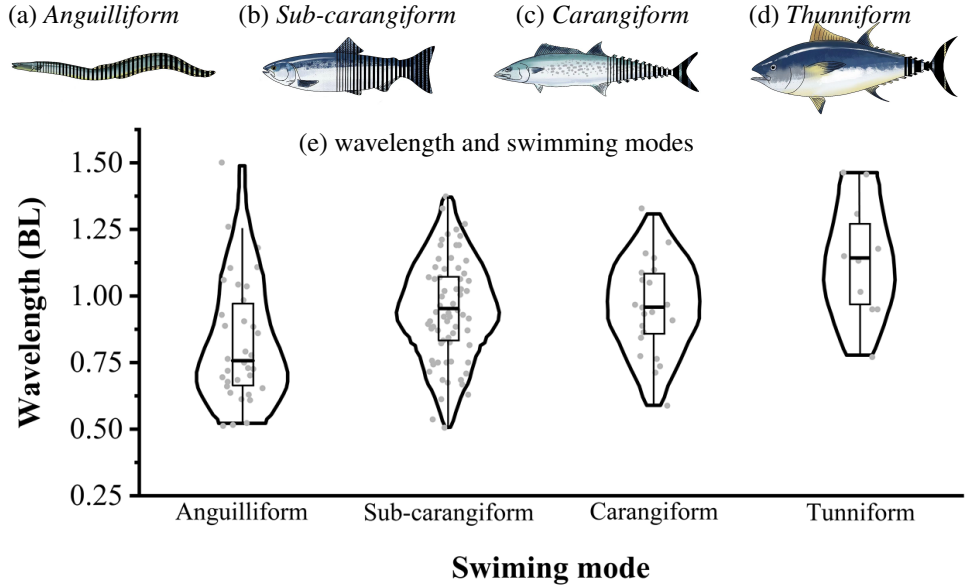

		\centering
		\addlabeltrim{0.8}{4_swimming_modes}{
			(\counterdd) \textit{Anguilliform} \ \ \ \ \ \ \ \   
			(\counterdd) \textit{Sub-carangiform} \ \ \ \ \ \ \ \ 
			(\counterdd) \textit{Carangiform} \ \ \ \ \ \ \ \ 
			(\counterdd) \textit{Thunniform}
		}
		\addlabelintro{0.8}{intro-swimmingform-wavelength-distribution-steady-swimming}{(\counterdd) wavelength and swimming modes}
		\caption{
			A classification of four most common BCF swimming modes (a) \textit{Anguilliform} with body undulation (e.g., eel), (b) \textit{Subcarangiform} with body undulation and caudal fin pitching (e.g., salmonid), (c) \textit{Carangiform} with minor body undulation and caudal fin pitching (e.g., mackerel), and (d) \textit{Thunniform} primarily relying on caudal fin pitching (e.g., tuna). The shaded area indicates the body regions involved in generating thrust, redrawn from figures by \cite{Lindsey1978} and \cite{Sfakiotakis1999}. (e) the distribution of wavelengths for these swimming modes. Regardless of the specific body-caudal fin sub-types, wavelengths range from 0.5 to 2 times the body length \citep{Santo2021} at steady swimming conditions.}
		\label{fig:4modes}
	\end{figure*}

	Numerical studies in both 2D and 3D have examined the effect of wavelength on single swimmers. \cite{chao2022hydrodynamic} investigated a wide range of parameters and identified seven wake structures, inspiring our parameter selection. \cite{Khalid2020} found that optimal hydrodynamic performance does not depend on specific wavelengths in natural swimmers. Several studies explored the influence of wavelength on thrust and propulsive efficiency \citep{Thekkethil2017, Thekkethil2018, Thekkethil2020, Gupta2021}. Additionally, \cite{Carling1998} studied a 2D anguilliform swimmer, while \cite{Khalid2021} focused on the advantages of shorter wavelengths for anguilliform swimmers. In 3D simulations, \cite{Borazjani2008, Borazjani2009} examined the effects of wavelengths on carangiform and anguilliform swimmers, respectively. To manage computational costs while still revealing fundamental patterns \citep{chao2022hydrodynamic}, we employ a 2D model in our current study.

	Wavelength significantly affects fish swimming during linear acceleration and speed \citep{Schwalbe2019}. Existing biological research has primarily focused on individual accelerating fish, with studies revealing differences in undulation kinematics between acceleration and steady swimming \citep{Schwalbe2019}. For bluegill sunfish, the body wavelength decreases during acceleration but increases with swimming speed, ranging from 0.75BL to 0.9BL at different acceleration levels \citep{Schwalbe2019}. In investigations of linear acceleration, tail-beat frequency has been found to have a greater impact on swimming speed and acceleration than amplitude \citep{Akanyeti2017}. Tail-beat amplitude remains constant during both steady swimming and acceleration \citep{Akanyeti2017}. Studies on anguilliform swimmers have shown a significant increase in both body wavelength and tailbeat frequency with steady swimming speed \citep{Tytell2004b}.
	
	Side-by-side fish schooling in steady swimming scenarios has been studied with fixed wavelengths. These studies have investigated the phalanx formation preference of fish at high steady-swimming speeds \citep{Ashraf2017}, as well as the maximum speed and efficiency achieved by side-by-side robotic fish in in-phase and anti-phase conditions \citep{Li2021}. Previous research on side-by-side and anti-phase pitching foils has demonstrated their ability to generate higher thrust with comparable efficiency to a single swimmer \citep{Dewey2014a, Huera-Huarte2018, Gungor2021, Yucel2022}. The influence of lateral distance on schooling performance has been explored, with studies indicating that a lateral distance of 0.33BL allows sufficient interaction between swimmers \citep{Li2020, Shrivastava2017, Wei2022}.

	In our previous studies, we have investigated the general effects of wavelengths on fish schooling \citep{lin2022swimming} and conducted detailed study regarding \textit{anti-phase} side-by-side scenarios \citep{lin2022wavelength}. However, \textit{in-phase} schooling is not yet systematically examined. In-phase schooling was discovered to maximise locomotion efficiency at a fixed wavelength \citep{Li2021}. How wavelength affects in-phase schooling at side-by-side conditions remains a question.
	So, in the present study, we continue to explore how fish-body wavelength affects the side-by-side schooling members undulating \textit{in-phase}.

	\FloatBarrier
	\clearpage
	
	\section{Methodology}
	In this section, we outline the methodology employed in the present study. The problem setup for schooling swimmers, encompassing geometry, kinematic equations, and non-dimensional analysis, is described in \Cref{sec_problem_setup}. Additionally, \Cref{sec_Comput_method} covers the computational method utilized to implement the problem setup.
	
	\subsection{Problem setup}
	\label{sec_problem_setup}
	
\begin{figure}
		\centering
		\includegraphics[width=1\linewidth]{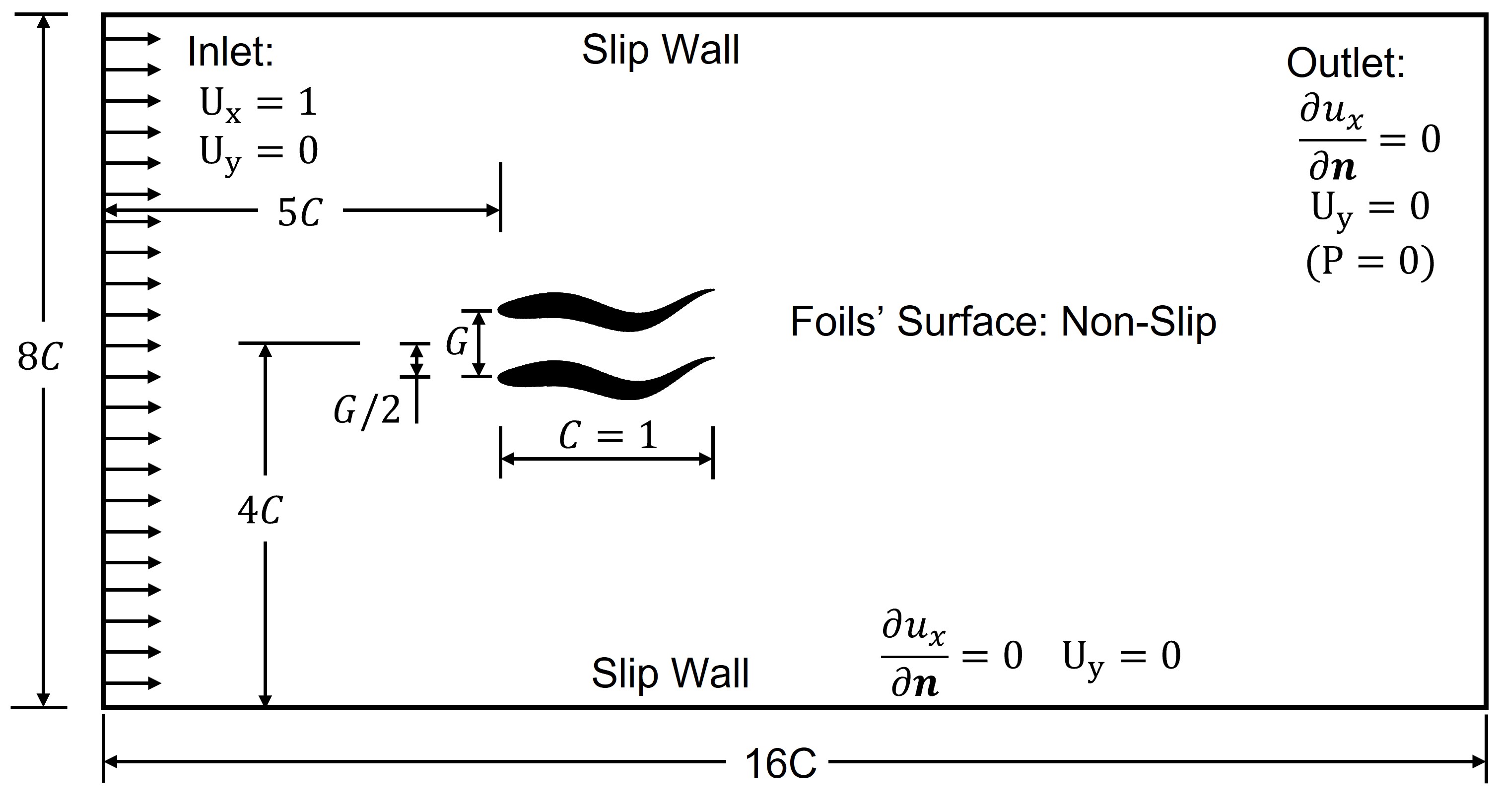}
		\caption{The schematic representation of the problem setup is depicted, illustrating two undulating tethered hydrofoils arranged in a phalanx (side-by-side) configuration, swimming \textit{in phase} with a fixed gap distance of $ G = 0.33 $.}
		\label{fig:problemsetup}
	\end{figure}

	The problem setup of the current study, illustrated in \Cref{fig:problemsetup}, focuses on the representation of accelerated fish schooling using a two-dimensional configuration with two wavy hydrofoils undulating side-by-side. This 2D configuration is appropriate for describing the laminar flow regime of interest, where $ Re \leq 2000 $ \citep{Gazzola2014, chao2022hydrodynamic}. The swimmers are tethered under prescribed flow to emulate instantaneous accelerating condition. This representation has been justified by \cite{Akanyeti2017} and our previous works \citep{lin2022swimming,lin2023wavelength}.

	For this study, we simplify the fish body as a 2D NACA0012 hydrofoil, which is widely utilised in examining bio-propulsion involving pitching \citep{Moriche2016} and undulating hydrofoils \citep{Thekkethil2017}.
	The geometry of a NACA0012 hydrofoil bears resemblance to that of a \textit{carangiform} or \textit{subcarangiform} swimmer.
	NACA hydrofoils, particularly the NACA0012 foil, are commonly employed as representative swimmer geometries \citep{Deng2022, Deng2015, Deng2016, Shao2010, Deng2007, pan2022effects, yu2021scaling}, enabling comparisons with previous investigations.
	To focus on essential variables associated with fish schooling, the two foils are positioned side-by-side, undulating in-phase. This is one of the typical scenarios observed in the nature \citep{Ashraf2017}.
	The swimmers' kinematics is described by the travelling wave equations in a non-dimensional form:
	
	\begin{equation}\label{equ:fish_wave_leader}
		Y_{\rm 1} = Y_{\rm 2} = A_{\rm max} X \sin \left[ 2\pi \left( \frac{X}{\lam} - \frac{St }{2 A_{\rm max}} t \right) \right]
	\end{equation}
	
	This configuration, which is commonly used \citep{Thekkethil2018}, is selected here for the purpose of convenient comparison. To provide a comprehensive understanding, the variables are defined as follows: $ Y_{\rm i} = Y'{\rm i} / C' $ represents the centre-line lateral displacement of each hydrofoil; $ i = 1 $ corresponds to the top swimmer, while $ i = 2 $ corresponds to the bottom swimmer; $ X = X' / C' $ indicates the streamwise position on the centreline. $ t = t' \uinf /C' $ represents the non-dimensional time, where $ \uinf $ denotes the free-stream velocity. $ A{\rm max} = A'{\rm max}/C' $ is the non-dimensional tail tip amplitude, while $ a{\rm max} $ represents the dimensional tail amplitude. $ \lam = \lam'/C' $ denotes the non-dimensional wavelength, where $ \lam' $ corresponds to the dimensional foil undulating wavelength. The Strouhal number is defined as $ St = 2f'A'_{\rm max} / \uinf $, with $ f' $ being the dimensional undulating frequency. In the notation used, dashed letters represent dimensional parameters.

	Furthermore, \Cref{tablecasegroups} presents non-dimensional groups and the explored parameter range for a specific case. Here, $ \rho' $ denotes fluid density, $ f' $ represents the undulating frequency, and $ \mu' $ stands for dynamic viscosity. In summary, the current investigation focuses on three variables: Reynolds number $ \re $ ranging from 1000 to 2000, Strouhal number $ St $ varying between 0.2 and 0.7, and non-dimensional wavelength $ \lam $ spanning from 0.5 to 2.0.

	\begin{ruledtabular}
		\begin{table*}[thb]
			\caption{Non-dimensional input parameters and the involved range of value}
			\centering
			\label{tablecasegroups}
			\begin{tabular}{l c c c}
				Wavelength  & $ \lam $ &{  $ {\lambda'}/{C'} $} & $ 0.5 - 2 $ \\
				
				Strouhal number  & $ St $ &{ $ {2 f' a'_{\rm max}}/{u'_{\rm \infty}} $} & $ 0.2 - 0.7 $ \\
				
				Reynolds number & $ \re $ &{ $ { \rho' u'_{\rm \infty} C'}/{\mu'} $} & $ 1000 - 2000 $ \\

			\end{tabular}
		\end{table*}
	\end{ruledtabular}
	
	The swimming performance metrics are presented in Table \ref{tab:output_para}. Thrust $ \bar{C}_{\rm T,i} $ is a key factor in acceleration, while net propulsive efficiency $ \eta_{\rm i} $ assesses the conversion of input energy into net thrust for acceleration. Analysing the vorticity field $ \bm{\omega} $ is essential for evaluating flow pattern and stealth capabilities. Table \ref{tab:output_para} provides the definitions of the variables used, where $ F_{\rm T,i} $ represents the net thrust on the hydrofoils and $ \bm{u} $ represents the fluid velocity.

	\begin{ruledtabular}
		\begin{table*}%
			\caption{Non-dimensional output parameters for swimming performance}
			\centering
			\label{tab:output_para}
			\begin{tabular}{l c c c}
				
				Cycle-averaged {net} thrust coefficient & $ \bar{C}_{\rm T,i} $ & $ \frac{1}{T} \int_{\rm t}^{t+T} C_T \diff t =  \frac{1}{T} \int_{\rm t}^{t+T} {2F_{\rm T,i}}/{\rho u^2_{\rm \infty} C} \diff t  $  \\

				Net propulsive efficiency & $ \eta_{\rm i} $ &{ $ {P_{\rm out,i}}/{P_{\rm in,i}} = {\bar{C}_T}/{\bar{C}_P} %
					$  }  \\

				Fluid vorticity field & $ \bm{\omega} $ &{ $ \nabla \times \bm{u} $  }  \\
				
			\end{tabular}
		\end{table*}
	\end{ruledtabular}
	
	\subsection{Immersed boundary method}
	\label{sec_Comput_method}

	In this study, a modified version \citep{lin2022swimming,lin2023wavelength} of the ConstraintIB module \citep{Bhalla2013,Griffith2020} within the IBAMR software framework \citep{griffith2013ibamr} was employed for simulation. IBAMR is an open-source immersed boundary method simulation software that relies on advanced libraries such as SAMRAI \citep{Hornung2002,Hornung2006}, PETSc \citep{Balay1997,Balay2010,balay2001petsc}, hypre \citep{falgout2010hypre,Balay1997}, and libmesh \citep{Kirk2006}. The selection of this software was based on its capability for adaptive mesh refinement of the Eulerian background mesh, ensuring computational efficiency and adequate accuracy. The ConstraintIB method has undergone extensive validation \citep{Bhalla2013,Bhalla2013a,bhalla2014fully,Nangia2017,Nangia2019,Griffith2020,Bhalla2020}, including validation of the customized version used in this study \citep{lin2022swimming}. Since the maximum Reynolds number in the present study, which is below 2000, differs from a previous study \citep{lin2022swimming} with $ Re = 5000 $, the same mesh refinement and time step settings that were verified for mesh independence were adopted. Each numerical simulation was conducted over 20 cycles of undulation.
	
	In the current investigation, the immersed boundary (IB) methodology employed is elucidated further here. The fluid is described using an Eulerian framework, while the deforming structure is articulated through a Lagrangian framework. A notable merit of this method lies in its computational efficiency, notably in bypassing the resource-intensive remeshing steps found in alternative techniques such as the finite element method. The formulation utilized is presented below:
	
	\begin{equation}\label{eq:IB1}
		\rho \left( \frac{\partial \bm{u}(\bm{x},t) }{\partial t} + \bm{u}(\bm{x},t) \cdot \nabla \bm{u}(\bm{x},t)\right) = -\nabla p (\bm{x},t) + \mu \nabla^2 \bm{u}(\bm{x},t) + \bm{f}(\bm{x},t)    
	\end{equation}
	\begin{equation}\label{eq:IB2}
		\nabla \cdot \bm{u}(\bm{x},t) = 0
	\end{equation}
	\begin{equation}\label{eq:IB3}
		\bm{f}(\bm{x},t) = \int_{U} \bm{F}(\bm{X},t) \delta(\bm{x} - \bm{\chi}(\bm{X},t)) \diff \bm{X}
	\end{equation}
	\begin{equation}\label{eq:IB4}
		\frac{\partial \bm{\chi} (\bm{X},t)}{\partial t} = \int_{\Omega} \bm{u}(\bm{x},t) \delta(\bm{x} - \bm{\chi}(\bm{X},t)) \diff \bm{x}
	\end{equation}
	
	In this context, $ \bm{x} = (x,y) \in \Omega $ denotes fixed physical Cartesian coordinates, where $ \Omega $ represents the physical domain occupied by the fluid and the immersed structure. $ \bm{X} = (X,Y) \in U $ designates Lagrangian solid structure coordinates, with $ U $ being the Lagrangian coordinate domain. The mapping from Lagrangian structure coordinates to the physical domain position of point $ \bm{X} $ for all time $ t $ is articulated as $ \bm{\chi}(\bm{X},t) = ( \chi_x(\bm{X},t), \chi_y(\bm{X},t) ) \in \Omega $. Put differently, $ \chi(U,t) \subset \Omega $ illustrates the physical region occupied by the solid structure at time $ t $. $ \bm{u}(\bm{x},t) $ symbolises the Eulerian fluid velocity field, while $ p(\bm{b},t) $ signifies the Eulerian pressure field. $ \rho $ represents the fluid density, and $ \mu $ denotes the incompressible fluid dynamic viscosity. $ \bm{f}(\bm{x},t) $ and $ \bm{F}(\bm{X},t) $ are the Eulerian and Lagrangian force densities, respectively. $ \delta (\bm{x}) $ is the Dirac delta function. Further information regarding the constrained IB formulation and discretisation procedure can be explored in preceding publications \citep{Bhalla2013,Griffith2020}.

	\FloatBarrier
	\clearpage
	
	\section{Results and discussion}
	\label{sec_result}
	
	In the present paper, we simulated $\caseNum$ cases in total, with wavelength, Strouhal number, and Reynolds number ranging as $ \lam = 0.5 - 2.0 $, $ \str = 0.2 - 0.7 $, and $ \re = 1000 - 2000 $. In the following sections, we will examine the corresponding variation of thrust force in \Cref{sec_ctm}, net propulsive efficiency in \Cref{sec_effi}, and the flow structures in \Cref{sec_flow}, with an attempt to summarise the general rules.
	
	\FloatBarrier
	
	\subsection{Net thrust $ \ctm $}
	\label{sec_ctm}
	How is the net thrust, which is proportional to the swimmer's acceleration, affected by $ \lam $, $ \re $ and $ \str $? The  answer can be summarised as the following empirical equation produced by the symbolic regression tool PySR \citep{pysr} and based on the simulation results as seen in \Cref{fig:heat_map_ctm}:	
	
	\begin{equation}\label{equ:SR_predict}
		\ctm = 1.3 \str^{3.30} \lam - \re^{-0.37} \lam^{-0.95} - 6.13 \re^{-0.60} %
	\end{equation}
	
	It should be noted that, due to the nature of in-phase schooling, the thrust force propelling the two swimmers are essentially identical for both swimmers. So in the present study, the net thrust is represented by $ \ctm = (C_{\rm 1} + C_{\rm 2}) / 2 $.
	
	The equation \Cref{equ:SR_predict} indicates a complex interplay among wavelength, Strouhal number, and the Reynolds number. The high exponent $ \str^{3.30} $ indicates the strong influence of the Strouhal number. 
	
	Here, for the convenience of comparison, we also present the equation describing the thrust force of anti-phase schooling as \Cref{equ:SR_predict_anti_phase} from \citep{lin2023wavelength} and single swimmer scenarios as \Cref{equ:chao_single_ctm} from \citep{chao2022hydrodynamic}.
	\begin{equation}\label{equ:SR_predict_anti_phase}
		\ctm = \re^{0.17} \str^{2.03} \lam^{1.23} - 0.26 \re^{0.19} \str^{1.00} \lam^{0.10} - 6.13 \re^{-0.6}
	\end{equation}
	
	\begin{equation}\label{equ:chao_single_ctm}
		\ctms = 0.36 \re^{0.208} \str^{3} \lam  - 6.13 \re^{-0.6}
	\end{equation}
	
	Comparing these equations, we can observe a few interesting facts.
	First, at in-phase condition, the Strouhal number $ \str^{3.3} $ is more dominant compared with the anti-phase scenario $ \str^{2.03} $, slightly more impactful than a single swimmer $ \str^3 $.
	Second, the wavelength at in-phase schooling, \ie $ \lam^1.0 $, is less influential compared with being anti-phase $ \lam^{1.23} $.
	In general, it is clear that, at in-phase schooling, the effects of these parameters behave more similarly to single-swimmer condition, rather than anti-phase schooling. This indicates that the underlying mechanism between the in-phase schooling and single-swimmer condition can be similar. For example, in a latter discussion of flow structure \Cref{sec_flow}, we found an interesting flow pattern of "2S-merge", where the shed vortices from two schooling swimmers merge into a single vortex street as if shed from a single swimmer.
	
	A more intuitive understanding of net thrust varition can be seen in the heat maps in \Cref{fig:heat_map_ctm}, where the thrust generally increases with wavelength and Strouhal number. Also, as seen in \Cref{fig:heat_map_ctm}f, the formula from the symbolic regression \Cref{equ:SR_predict} matches the simulation results well with $ R^2 = 0.986 $.
	
	Furthermore, we also compare the thrust force produced by the minimal school with that of a single swimmer, by the \textit{schooling thrust amplification factor} $ \ctm/\ctms $, as seen in \Cref{fig:heat_map_compare_12_ctm}. It is clear that \textit{in-phase schooling} always produces thrust force that is smaller than a single swimmer with $ \ctm/\ctms < 0.66 < 1 $, which is in drastic contrast with \textit{anti-phase schooling} \citep{lin2023wavelength} where $ \ctm/\ctms \approx 5 $ can be reached in the same parametric space. So this indicates that in-phase swimming with phalanx formation does not allow the school to accelerate faster than a single swimmer. 
	
	\newcounter{testa}
	\setcounter{testa}{0}
	\newcommand\countera{\stepcounter{testa}\alph{testa}}
	\newcommand{\addlabele}[3]{%
		\begin{tikzpicture}
			\node[anchor=south west,inner sep=0] (image) at (0,0) 
			{\includegraphics[width=#1\textwidth]{#2}};
			\begin{scope}[x={(image.south east)},y={(image.north west)}]
				\node[anchor=south west] at (0.15,0.90) {\footnotesize #3};
			\end{scope}
		\end{tikzpicture}%
	}
	\begin{figure}
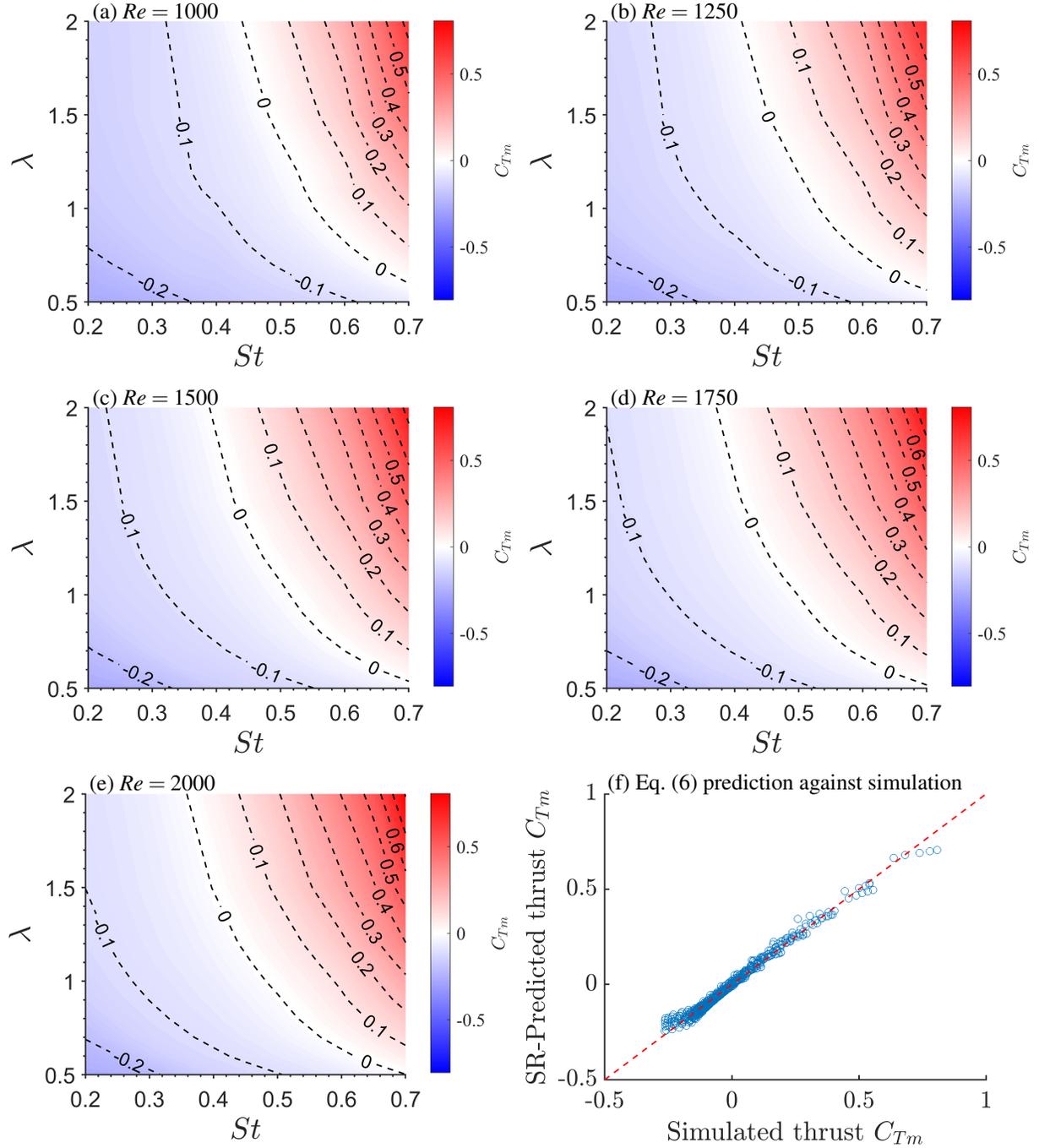

		\centering
		\setcounter{testa}{0}
		\addlabele{0.49}{{{Heat_map_Re_1000__CTm_St_lambda_ppi1_D0_G0d35}}}{(\countera) $ Re = 1000 $}
		\addlabele{0.49}{{{Heat_map_Re_1250__CTm_St_lambda_ppi1_D0_G0d35}}}{(\countera) $ Re = 1250 $}
		\addlabele{0.49}{{{Heat_map_Re_1500__CTm_St_lambda_ppi1_D0_G0d35}}}{(\countera) $ Re = 1500 $}
		\addlabele{0.49}{{{Heat_map_Re_1750__CTm_St_lambda_ppi1_D0_G0d35}}}{(\countera) $ Re = 1750 $}
		\addlabele{0.49}{{{Heat_map_Re_2000__CTm_St_lambda_ppi1_D0_G0d35}}}{(\countera) $ Re = 2000 $}
		\addlabele{0.49}{{{ctm_SR_prediction_accuracy_line}}}{(\countera) \Cref{equ:SR_predict} prediction against simulation}
		\setcounter{testa}{0}
		\caption{Heat map displaying the average net thrust, denoted as $ \ctm $, over Strouhal number range $ St = 0.2-0.7 $, wavelength interval $ \lam=0.5-2 $ and Reynolds numbers at 
			(a) $ Re = 1000 $
			(b) $ Re = 1250 $
			(c) $ Re = 1500 $
			(d) $ Re = 1750 $
			(e) $ Re = 2000 $.
			(f) Symbolic regression prediction accuracy comparing simulation results and \Cref{equ:SR_predict}, where $ R^2 = 0.986 $. The thrust on two swimmers is almost identical.
			The representation of forward acceleration, or positive thrust, is depicted in red. On the other hand, blue denotes deceleration, or negative thrust. Additionally, the contour line where $\ctm = 0$ symbolizes scenarios of zero net thrust, indicative of a constant swimming state.
		}
		\label{fig:heat_map_ctm}
	\end{figure}

\setcounter{testa}{0}
\renewcommand{\addlabele}[3]{%
	\begin{tikzpicture}
		\node[anchor=south west,inner sep=0] (image) at (0,0) 
		{\includegraphics[width=#1\textwidth]{#2}};
		\begin{scope}[x={(image.south east)},y={(image.north west)}]
			\node[anchor=south west] at (0.15,0.90) {\footnotesize #3};
		\end{scope}
	\end{tikzpicture}%
}
\newcommand{\widthb}{0.32}
\begin{figure}
	\centering
	\setcounter{testa}{0}
	\addlabele{\widthb}{{{compare__Heat_map_Re_1000__CTm_St_lambda_ppi1_D0_G0d35}}}{(\countera) $ Re = 1000 $}
	\addlabele{\widthb}{{{compare__Heat_map_Re_1250__CTm_St_lambda_ppi1_D0_G0d35}}}{(\countera) $ Re = 1250 $}
	\addlabele{\widthb}{{{compare__Heat_map_Re_1500__CTm_St_lambda_ppi1_D0_G0d35}}}{(\countera) $ Re = 1500 $}
	\addlabele{\widthb}{{{compare__Heat_map_Re_1750__CTm_St_lambda_ppi1_D0_G0d35}}}{(\countera) $ Re = 1750 $}
	\addlabele{\widthb}{{{compare__Heat_map_Re_2000__CTm_St_lambda_ppi1_D0_G0d35}}}{(\countera) $ Re = 2000 $}
	\addlabele{\widthb}{{{compare__Line_Re_vs_largest_thrust_Amplification}}}{(\countera) Maximum amplification}
	\setcounter{testa}{0}
	\caption{Heat map illustrating the factor of \textit{schooling thrust amplification}, represented as $ \ctm/\ctms $, across Strouhal numbers $ \str = 0.2-0.7 $, wavelengths $ \lam=0.5-2 $ and Reynolds numbers at
		(a) $ Re = 1000 $
		(b) $ Re = 1250 $
		(c) $ Re = 1500 $
		(d) $ Re = 1750 $
		(e) $ Re = 2000 $
		(f) Maximum thrust enhancement due to schooling.
		In this presentation, only cases where both $ \ctm > 0 $ and $ \ctms > 0 $ are considered; scenarios that do not involve acceleration are represented as zero.
		It is seen that, in the tested range of parameters, \textit{in-phase} schooling always results in less thrust power compared with \textit{anti-phase} scenarios \citep{lin2023wavelength} where the schooling thrust can reach several times for that of the single swimmer.
	}
	\label{fig:heat_map_compare_12_ctm}
\end{figure}

\FloatBarrier
	\subsection{Net propulsive efficiency $ \eta $ }
	\label{sec_effi}
	
	In this section, we discuss the variation of net propulsive efficiency $\eta$ in the tested range. It should be noted that here $\eta$ is defined as the average value of the efficiency from two schooling swimmers as $ \eta = (\eta_1 + \eta_2) / 2 $, because due to the nature of in-phase schooling, $\eta_1 \approx \eta_2 $ establishes for almost all involved cases. Also, $\eta = \bar{C}_{T,pair}/\bar{C}_P$ here is the net Froude efficiency to quantify the the conversion effectiveness of input power into net thrust.
	
	Overall, at in-phase schooling, the efficiency increases with Strouhal number, $\str$, whereas the optimal efficiency is obtained at wavelengths of $ \lam = 1.1 - 1.5 $, almost regardless of $\re $, as seen in \Cref{fig:heat_map_eta}. Also, it is seen that the optimal efficiency should be obtained at Strouhal number of $\str > 0.7$, beyond the current parametric range. Due to constraint of computational resources, we cannot explore beyond this range, which is chosen with relevance of fish swimming in nature, as previously discussed \citep{lin2023wavelength,lin2022swimming}. Nevertheless, these results of \textit{in-phase} schooling is already in good contrast with \textit{anti-phase} cases \citep{lin2023wavelength}. Comparing the in-phase and anti-phase schooling, the optimal wavelength is similarly in the range of $\lam = 1 - 1.5$; however, the Strouhal number required for optimal efficiency of in-phase schooling, $\str > 0.7$, is much higher than that of the anti-phase schooling at $ \str \approx 0.55 $. Also, it is interesting to note that the maximum efficiency linearly increases with $\re$, as demonstrated in \Cref{fig:heat_map_eta}f, being similar to anti-phase \citep{lin2023wavelength}.
	
	In addition, we also compare the efficiency of in-phase schooling with that of a single swimmer, by drawing heat maps for \textit{efficiency amplification factor}, $\etaamp$, as seen in \Cref{fig:heat_map_compare_12_eta}. Again, similar to discussion for thrust force in \Cref{sec_ctm}, the efficiency of in-phase schooling never outperforms that of a single swimmer with $\etaamp < 0.96 $, whereas for anti-phase schooling, it can reach many times of a single swimmer \citep{lin2023wavelength}.

	\newcommand{\widtha}{0.40}
	\begin{figure}
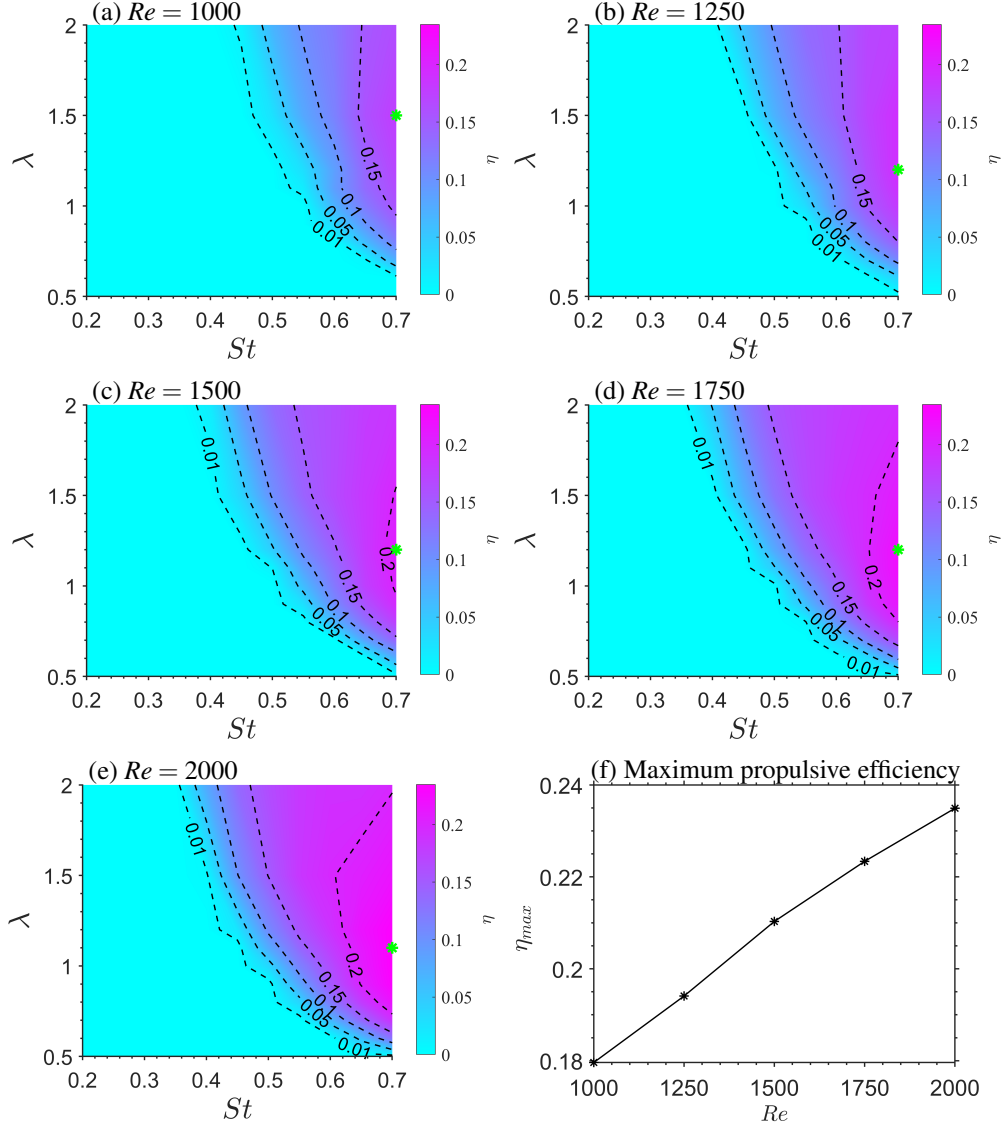

		\centering
		\setcounter{testa}{0}
		\addlabele{\widtha}{{{Heat_map_Re_1000__Effi_St_lambda_ppi1_D0_G0d35}}}{(\countera) $ Re = 1000 $}
		\addlabele{\widtha}{{{Heat_map_Re_1250__Effi_St_lambda_ppi1_D0_G0d35}}}{(\countera) $ Re = 1250 $}
		\addlabele{\widtha}{{{Heat_map_Re_1500__Effi_St_lambda_ppi1_D0_G0d35}}}{(\countera) $ Re = 1500 $}
		\addlabele{\widtha}{{{Heat_map_Re_1750__Effi_St_lambda_ppi1_D0_G0d35}}}{(\countera) $ Re = 1750 $}
		\addlabele{\widtha}{{{Heat_map_Re_2000__Effi_St_lambda_ppi1_D0_G0d35}}}{(\countera) $ Re = 2000 $}
		\addlabele{\widtha}{{{Line_Re_vs_Optimal_Effi}}}{(\countera) Maximum propulsive efficiency}
		\setcounter{testa}{0}
		\caption{Heat map depicting \textit{net propulsive efficiency}, denoted by $ \eta $, across Strouhal numbers $ \str = 0.2-0.7 $, wavelength range $ \lam=0.5-2 $, and varying Reynolds numbers at
			(a) $ Re = 1000 $
			(b) $ Re = 1250 $
			(c) $ Re = 1500 $
			(d) $ Re = 1750 $
			(e) $ Re = 2000 $.
			(f) Peak net propulsive efficiency, represented as $ \eta_{\rm max} $, corresponding to each specified Reynolds number.
			Owing to the side-by-side configuration, the net propulsive efficiency is the same for each individual swimmer as well as for the two swimmers collectively.
			The highest efficiency is denoted by the green star marker; the maximum efficiency is obtained at $ St = 0.7 $ regardless of Reynolds number, whereas the corresponding optimal wavelength ranges from $ \lam = 1.1 - 1.5 $. The maximum propulsive efficiency, obtainable by adjusting Strouhal number and wavelength, linearly increases with Reynolds number from $ 18\% $ to $ 24\% $.
			The cases with negative thrust are drawn as zero.
			Compared with the anti-phase conditions \citep{lin2023wavelength}, the optimal strouhal number is higher; the optimal wavelength is roughly in the same range; the obtained maximum efficiency is generally lower. So while accelerating from low Reynolds number $ \re<2000 $, anti-phase is more advantageous than in-phase condition.
		}
		\label{fig:heat_map_eta}
	\end{figure}

	\setcounter{testa}{0}
	\renewcommand{\addlabele}[3]{%
		\begin{tikzpicture}
			\node[anchor=south west,inner sep=0] (image) at (0,0) 
			{\includegraphics[width=#1\textwidth]{#2}};
			\begin{scope}[x={(image.south east)},y={(image.north west)}]
				\node[anchor=south west] at (0.15,0.90) {\footnotesize #3};
			\end{scope}
		\end{tikzpicture}%
	}
	\renewcommand{\widthb}{0.32}
	\begin{figure}
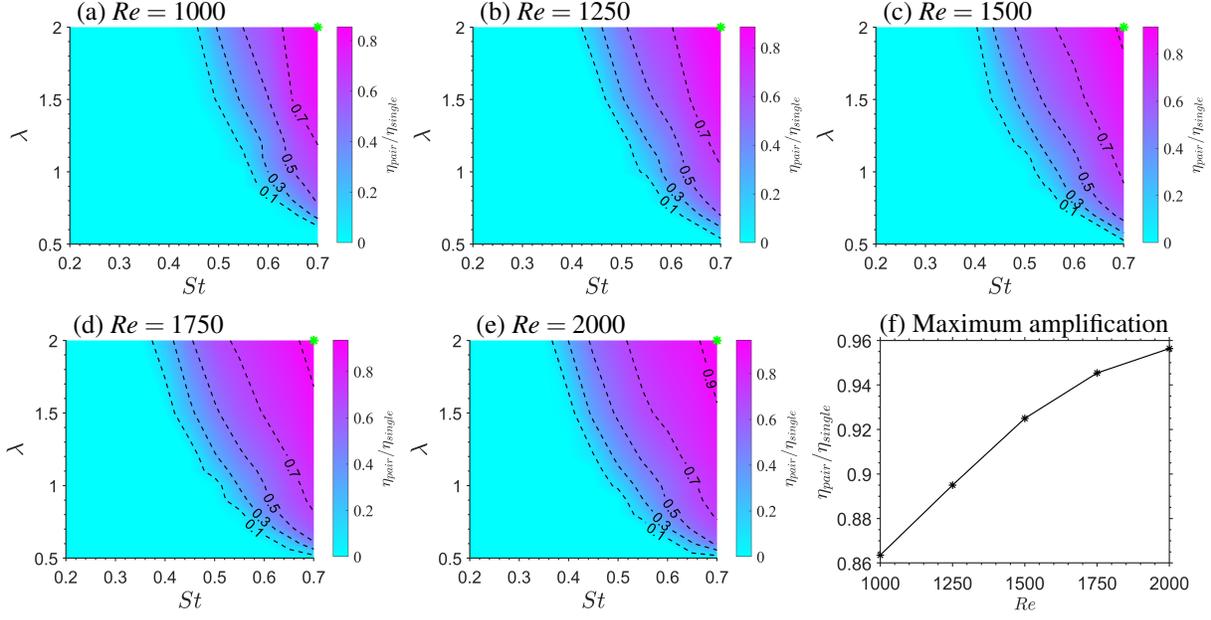

		\centering
		\setcounter{testa}{0}
		\addlabele{\widthb}{{{compare__Heat_map_Re_1000__Eta_St_lambda_ppi1_D0_G0d35}}}{(\countera) $ Re = 1000 $}
		\addlabele{\widthb}{{{compare__Heat_map_Re_1250__Eta_St_lambda_ppi1_D0_G0d35}}}{(\countera) $ Re = 1250 $}
		\addlabele{\widthb}{{{compare__Heat_map_Re_1500__Eta_St_lambda_ppi1_D0_G0d35}}}{(\countera) $ Re = 1500 $}
		\addlabele{\widthb}{{{compare__Heat_map_Re_1750__Eta_St_lambda_ppi1_D0_G0d35}}}{(\countera) $ Re = 1750 $}
		\addlabele{\widthb}{{{compare__Heat_map_Re_2000__Eta_St_lambda_ppi1_D0_G0d35}}}{(\countera) $ Re = 2000 $}
		\addlabele{\widthb}{{{compare__Line_Re_vs_largest_Efficiency_Amplification}}}{(\countera) Maximum amplification}
		\setcounter{testa}{0}
		\caption{Heat map for \textit{in-phase schooling efficiency amplification factor} $ \eta_{\rm pair} / \eta_{\rm single} $ at Strouhal number $ \str = 0.2-0.7 $, wavelength $ \lam=0.5-2 $ and Reynolds numbers at 
			(a) $ Re = 1000 $
			(b) $ Re = 1250 $
			(c) $ Re = 1500 $
			(d) $ Re = 1750 $
			(e) $ Re = 2000 $
			(f) Peak efficiency enhancement at each $ \re $ as a result of schooling.
			In this section, we present only those results where both $ \ctm > 0 $ and $ \ctms > 0 $ are observed, while scenarios involving no acceleration are depicted as zero.
			This map demonstrates how in-phase schooling reduces the propulsive efficiency of the swimmers.
			Across all the tested cases, amplifcation factor is less than one $ \etamp < 1 $, meaning that schooling swimmers always produce lower propulsive efficiency than a single swimmer. This is in drastic contrast with the amplification factors for the \textit{anti-phase} cases, where the factor can reach $ \etamp \approx 5 $ \citep{lin2023wavelength}.
			The green marker indicates the position corresponding to the maximum efficiency amplification at each Reynolds number.
			The optimal efficiency amplification, obtainable by adjusting Strouhal number and wavelength, increases with Reynolds number, differing from the anti-phase results \citep{lin2023wavelength}.
		}
		\label{fig:heat_map_compare_12_eta}
	\end{figure}

	\FloatBarrier
	\subsection{Flow structure maps}
	\label{sec_flow}
	
	In this section, we present the unique flow structures while the two swimmers acclerate side-by-side. Five distinct types of flow structures are identified, as seen in \Cref{fig:example_flow_struct}. In general, the flow structures are periodical without symmetry breaking. This indicates that, in the present parametric space relevant to most swimming fish species, symmetry breaking does not occur. Also, the skewing of streaming direction only takes place with type (d) 1P-skew, while other patterns' streaming direction is parallel to the accelerating direction. This indicates that the school of two swimmers will only generate a skewed wake at type (d). The characteristics of each flow structure types is summarised in the following list:
	
	\begin{enumerate}[label=(\alph*)]
		\item 2S-outer: in each cycle, two single vortices are shed from each swimmer, moving downstream at outer sides of the swimmers.
		\item 2P-quasi: two pairs of vortices shed per cycle from each swimmer, forming two stable vortex streets.
		\item 2S-merge: two single vortices per cycle from each swimmer, merging to form a single vortex street in the wake.
		\item 1P-skew: two pairs of vortices shed per cycle from each swimmer, but after shedding, they immediately merge into one pair of vortices, forming a skewed vortex street.
		\item 2P-diverge: two pairs of vortices shed per cycle from each swimmer, forming two vortex streets that repels each other, resulting in a diverging pattern.
	\end{enumerate}

	In addition, we draw the maps to demonstrate the distribution of flow pattern types in the tested parametric space with $ \re = 1000 - 2000 $, $ \lam = 0.5 - 2.0 $, and $ \str = 0.2 - 0.7 $, as seen in \Cref{fig:map_flow_struct}. Being similar to anti-phase scenarios \citep{lin2023wavelength}, the distribution is not significantly affected by Reynolds number at $ \re = 1000 - 2000 $, whereas the wavelength and strouhal number is more impactful. More specifically, (a) 2S-outer (b) 2P-quasi and (e) 2P-diverge all take place at roughly $ 0.5 < \lam < 1.2 $. At the bottom-left corner of \Cref{fig:map_flow_struct}, the (a) 2S-outer is observed at $ 0.2 < \str < 0.4 $; with the increase of Strouhal number, transition to 2P-quasi happens at $ \str \approx 0.3 $, and then to (e) 2P-diverge at $  \str \approx 0.5 $. So it is clear that the increase of Strouhal number, which can also be understood as non-dimensional frequency, causes the weak vortices at (a) 2S-outer to become a clear formation at (b) 2P-quasi; and then being strong enough for the top and bottom vortex streets to repel each other, as seen in (e) 2P-diverge of \Cref{fig:map_flow_struct}.
	Apart from that, (c) 2S-merge, (d) 1P-skew and (e) 2P-diverge are dicovered at a higher wavelength $ 1.2 < \lam < 2.0 $. Starting from the top-left corner of the regime with large wavelength $ 1.2 < \lam < 2 $ and low Strouhal number $ 0.2 < \str < 0.4 $, (c) 2S-merge is identified. As $ \str $ increases above $ 0.4 $, the merged vortices change into a skewed street of vortex dipoles as seen in (d) 1P-skew; and then at $ \str > 0.5 $, the pattern is transitioned to (e) 2P-diverge as well. So here at the regime of relatively high wavelength, with the increase of strouhal number, the flow pattern changes from the merged wake to a skewed vortex street and then to 2 diverged vortex streets.
	
	In addition, considering the results regarding thrust and efficiency in previous sections, we found that the highest thrust and propulsive efficiency are both discovered in (e) 2P-diverge regime with high Strouhal numbers.
	
	\renewcommand{\addlabele}[3]{%
		\begin{tikzpicture}
			\node[anchor=south west,inner sep=0] (image) at (0,0) 
			{\includegraphics[width=#1\textwidth, trim={3.5cm 0cm 3cm 0cm},clip]{#2}};
			\begin{scope}[x={(image.south east)},y={(image.north west)}]
				\node[anchor=south west] at (0.15,0.90) {\footnotesize #3};
			\end{scope}
		\end{tikzpicture}%
	}
	\renewcommand{\widtha}{0.32}
	\begin{figure}
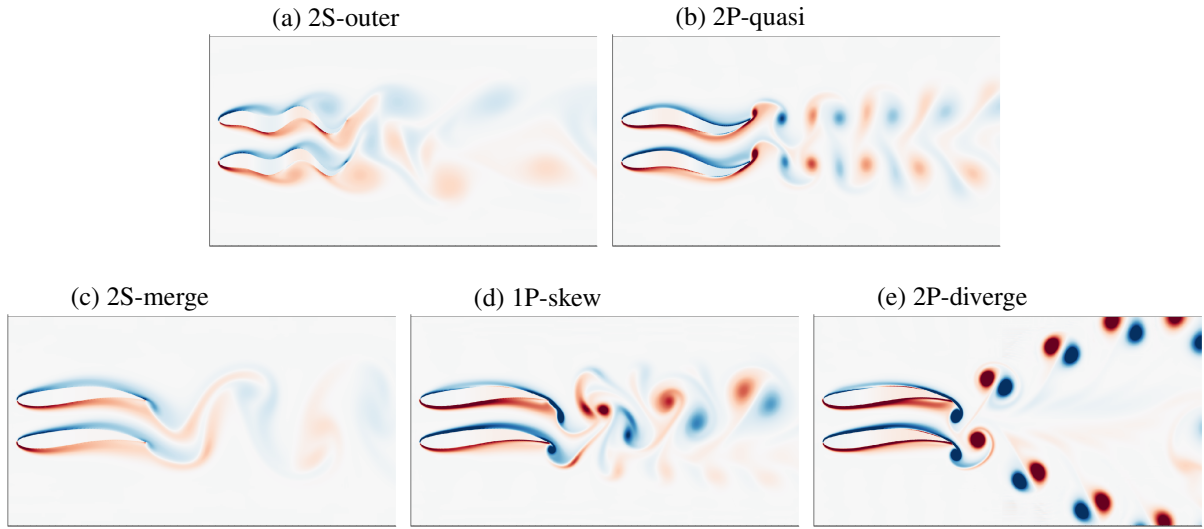

		\centering
		\setcounter{testa}{0}
		\addlabele{\widtha}{{{example_Flow_structure_Re_1000_St_0.2_Lambda_0.5}}}{(\countera) 2S-outer}
		\addlabele{\widtha}{{{example_Flow_structure_Re_1000_St_0.45_Lambda_1}}}{(\countera) 2P-quasi}
		
		\addlabele{\widtha}{{{example_Flow_structure_Re_1000_St_0.2_Lambda_2}}}{(\countera) 2S-merge}
		\addlabele{\widtha}{{{example_Flow_structure_Re_1000_St_0.45_Lambda_2}}}{(\countera) 1P-skew}
		\addlabele{\widtha}{{{example_Flow_structure_Re_1000_St_0.7_Lambda_2}}}{(\countera) 2P-diverge}
		
		\setcounter{testa}{0}
		\caption{Five distinc flow structures are identified as:
			(a) 2S-outer
			(b) 2P-quasi
			(c) 2S-merge
			(d) 1P-skew
			(e) 2P-diverge
			Here, the red colour denotes positive vorticity (counter-clockwise) with the blue colour representing the negative vorticity (clockwise).
		}
		\label{fig:example_flow_struct}
	\end{figure}
	\setcounter{testa}{0}

	\renewcommand{\addlabele}[3]{%
		\begin{tikzpicture}
		\node[anchor=south west,inner sep=0] (image) at (0,0) 
		{\includegraphics[width=#1\textwidth, trim={0.75cm 0cm 1cm 0cm},clip]{#2}};
		\begin{scope}[x={(image.south east)},y={(image.north west)}]
		\node[anchor=south west] at (0.15,0.90) {\footnotesize #3};
		\end{scope}
		\end{tikzpicture}%
	}
	\newcommand{\addlabelcutlegend}[2]{%
		\begin{tikzpicture}
		\node[anchor=south west,inner sep=0] (image) at (0,0) 
		{\includegraphics[width=#1\linewidth, trim={9.5cm 3cm 0.5cm 2.75cm},clip]{#2}};	%
		\begin{scope}[x={(image.south east)},y={(image.north west)}]
		\end{scope}
		\end{tikzpicture}%
	}
	\renewcommand{\widtha}{0.32}
	\begin{figure}
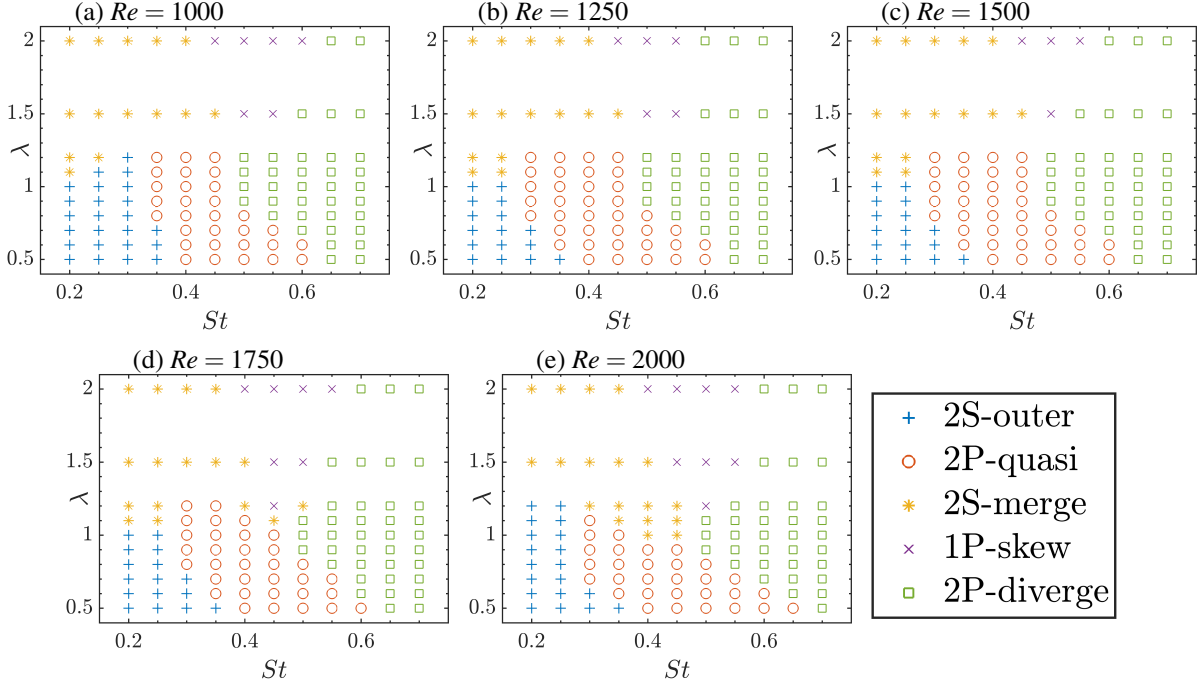

		\centering
		\setcounter{testa}{0}
		\addlabele{\widtha}{{{mapping_Flow_structure_Re_1000}}}{(\countera) $ Re = 1000 $}
		\addlabele{\widtha}{{{mapping_Flow_structure_Re_1250}}}{(\countera) $ Re = 1250 $}
		\addlabele{\widtha}{{{mapping_Flow_structure_Re_1500}}}{(\countera) $ Re = 1500 $}
		\addlabele{\widtha}{{{mapping_Flow_structure_Re_1750}}}{(\countera) $ Re = 1750 $}
		\addlabele{\widtha}{{{mapping_Flow_structure_Re_2000}}}{(\countera) $ Re = 2000 $}
		\addlabelcutlegend{0.22}{{{legend_for_mapping_Flow_structure}}}
		\setcounter{testa}{0}
		\caption{Distribution of flow structures across Strouhal numbers $ St = 0.2-0.7 $, wavelengths $ \lam=0.5-2 $, and various Reynolds numbers at
			(a) $ Re = 1000 $
			(b) $ Re = 1250 $
			(c) $ Re = 1500 $
			(d) $ Re = 1750 $
			(e) $ Re = 2000 $.
			The corresponding classification is shown in \Cref{fig:example_flow_struct}. Overall, the variation of reynolds number from 1000 to 2000 does not significantly alter the flow pattern distribution, whereas Strouhal number and wavelength play a greater role.
		}
		\label{fig:map_flow_struct}
	\end{figure}
	\setcounter{testa}{0}

	\FloatBarrier
	\section{Conclusions}
	In conclusion, in-phase schooling is not advantageous for linear acceleration compared with anti-phase or even with a single swimmer, at least in the current parametric space of $\str = 0.2 - 0.7$, $\lam = 0.5 - 2$, and $\re = 1000 - 2000$, considering acceleration-relevant metrics, \ie net thrust force (\Cref{sec_ctm}) and net propulsive efficiency (\Cref{sec_effi}). Five distinct types of flow structures are identified (\Cref{sec_flow}), without the observation of symmetry breaking. At certain parameters, the wake pattern is very similar to that from a single swimmer, as "2S-merge". Also, asymmetrical vortex street is observed as "1P-skew".
	More specifically, the net thrust and the net propulsive efficiency never surpasses that of a single swimmer, indicating impaired acceleration performance due to \textit{in-phase} schooling, not to mention comparing with the much more advantageous performance of \textit{anti-phase} schooling \cite{lin2023wavelength}. The phase difference, while simple to modify in engineering contexts, can significantly impact the linear acceleration of a minimal school. It also further indicates the importance of sensoring and coordinating with the nearby swimmers in a school during acceleration conditions.

	\begin{acknowledgments}
		This work was funded by
		China Postdoctoral Science Foundation (Grant No. 2021M691865)
		and by
		Science and Technology Major Project of Fujian Province in China (Grant No. 2021NZ033016).
		We appreciate the US National Science Foundation award OAC 1931368 (A.P.S.B) for supporting the IBAMR library.
		This work was also financially supported by the National Natural Science Foundation of China (Grant Nos. 12074323; 42106181), the Natural Science Foundation of Fujian Province of China (No. 2022J02003), the China National Postdoctoral Program for Innovative Talents (Grant No. BX2021168) and the Outstanding Postdoctoral Scholarship, State Key Laboratory of Marine Environmental Science at Xiamen University.
		
	\end{acknowledgments}
	
	\section*{Data Availability Statement}
	The data that support the findings of this study are available from the corresponding author upon reasonable request.
	
	\FloatBarrier
	\bibliography{library}%

\begin{thebibliography}{76}%
\makeatletter
\providecommand \@ifxundefined [1]{%
 \@ifx{#1\undefined}
}%
\providecommand \@ifnum [1]{%
 \ifnum #1\expandafter \@firstoftwo
 \else \expandafter \@secondoftwo
 \fi
}%
\providecommand \@ifx [1]{%
 \ifx #1\expandafter \@firstoftwo
 \else \expandafter \@secondoftwo
 \fi
}%
\providecommand \natexlab [1]{#1}%
\providecommand \enquote  [1]{``#1''}%
\providecommand \bibnamefont  [1]{#1}%
\providecommand \bibfnamefont [1]{#1}%
\providecommand \citenamefont [1]{#1}%
\providecommand \href@noop [0]{\@secondoftwo}%
\providecommand \href [0]{\begingroup \@sanitize@url \@href}%
\providecommand \@href[1]{\@@startlink{#1}\@@href}%
\providecommand \@@href[1]{\endgroup#1\@@endlink}%
\providecommand \@sanitize@url [0]{\catcode `\\12\catcode `\$12\catcode
  `\&12\catcode `\#12\catcode `\^12\catcode `\_12\catcode `\%12\relax}%
\providecommand \@@startlink[1]{}%
\providecommand \@@endlink[0]{}%
\providecommand \url  [0]{\begingroup\@sanitize@url \@url }%
\providecommand \@url [1]{\endgroup\@href {#1}{\urlprefix }}%
\providecommand \urlprefix  [0]{URL }%
\providecommand \Eprint [0]{\href }%
\providecommand \doibase [0]{https://doi.org/}%
\providecommand \selectlanguage [0]{\@gobble}%
\providecommand \bibinfo  [0]{\@secondoftwo}%
\providecommand \bibfield  [0]{\@secondoftwo}%
\providecommand \translation [1]{[#1]}%
\providecommand \BibitemOpen [0]{}%
\providecommand \bibitemStop [0]{}%
\providecommand \bibitemNoStop [0]{.\EOS\space}%
\providecommand \EOS [0]{\spacefactor3000\relax}%
\providecommand \BibitemShut  [1]{\csname bibitem#1\endcsname}%
\let\auto@bib@innerbib\@empty
\bibitem [{\citenamefont {Akanyeti}\ \emph {et~al.}(2017)\citenamefont
  {Akanyeti}, \citenamefont {Putney}, \citenamefont {Yanagitsuru},
  \citenamefont {Lauder}, \citenamefont {Stewart},\ and\ \citenamefont
  {Liao}}]{Akanyeti2017}%
  \BibitemOpen
  \bibfield  {author} {\bibinfo {author} {\bibnamefont {Akanyeti},
  \bibfnamefont {O.}}, \bibinfo {author} {\bibnamefont {Putney}, \bibfnamefont
  {J.}}, \bibinfo {author} {\bibnamefont {Yanagitsuru}, \bibfnamefont {Y.~R.}},
  \bibinfo {author} {\bibnamefont {Lauder}, \bibfnamefont {G.~V.}}, \bibinfo
  {author} {\bibnamefont {Stewart}, \bibfnamefont {W.~J.}}, and\ \bibinfo
  {author} {\bibnamefont {Liao}, \bibfnamefont {J.~C.}},\ }\bibfield  {title}
  {\enquote {\bibinfo {title} {{Accelerating fishes increase propulsive
  efficiency by modulating vortex ring geometry}},}\ }\href
  {https://doi.org/10.1073/pnas.1705968115} {\bibfield  {journal} {\bibinfo
  {journal} {Proceedings of the National Academy of Sciences of the United
  States of America}\ }\textbf {\bibinfo {volume} {114}} (\bibinfo {year}
  {2017}),\ 10.1073/pnas.1705968115}\BibitemShut {NoStop}%
\bibitem [{\citenamefont {Ashraf}\ \emph {et~al.}(2017)\citenamefont {Ashraf},
  \citenamefont {Bradshaw}, \citenamefont {Ha}, \citenamefont {Halloy},
  \citenamefont {Godoy-Diana},\ and\ \citenamefont {Thiria}}]{Ashraf2017}%
  \BibitemOpen
  \bibfield  {author} {\bibinfo {author} {\bibnamefont {Ashraf}, \bibfnamefont
  {I.}}, \bibinfo {author} {\bibnamefont {Bradshaw}, \bibfnamefont {H.}},
  \bibinfo {author} {\bibnamefont {Ha}, \bibfnamefont {T.~T.}}, \bibinfo
  {author} {\bibnamefont {Halloy}, \bibfnamefont {J.}}, \bibinfo {author}
  {\bibnamefont {Godoy-Diana}, \bibfnamefont {R.}}, and\ \bibinfo {author}
  {\bibnamefont {Thiria}, \bibfnamefont {B.}},\ }\bibfield  {title} {\enquote
  {\bibinfo {title} {{Simple phalanx pattern leads to energy saving in cohesive
  fish schooling}},}\ }\href {https://doi.org/10.1073/pnas.1706503114}
  {\bibfield  {journal} {\bibinfo  {journal} {Proceedings of the National
  Academy of Sciences of the United States of America}\ }\textbf {\bibinfo
  {volume} {114}} (\bibinfo {year} {2017}),\
  10.1073/pnas.1706503114}\BibitemShut {NoStop}%
\bibitem [{\citenamefont {Balay}\ \emph
  {et~al.}(2023{\natexlab{a}})\citenamefont {Balay}, \citenamefont {Abhyankar},
  \citenamefont {Adams}, \citenamefont {Benson}, \citenamefont {Brown},
  \citenamefont {Brune}, \citenamefont {Buschelman}, \citenamefont
  {Constantinescu}, \citenamefont {Dalcin}, \citenamefont {Dener},
  \citenamefont {Eijkhout}, \citenamefont {Faibussowitsch}, \citenamefont
  {Gropp}, \citenamefont {Hapla}, \citenamefont {Isaac}, \citenamefont
  {Jolivet}, \citenamefont {Karpeev}, \citenamefont {Kaushik}, \citenamefont
  {Knepley}, \citenamefont {Kong}, \citenamefont {Kruger}, \citenamefont {May},
  \citenamefont {McInnes}, \citenamefont {Mills}, \citenamefont {Mitchell},
  \citenamefont {Munson}, \citenamefont {Roman}, \citenamefont {Rupp},
  \citenamefont {Sanan}, \citenamefont {Sarich}, \citenamefont {Smith},
  \citenamefont {Zampini}, \citenamefont {Zhang}, \citenamefont {Zhang},\ and\
  \citenamefont {Zhang}}]{balay2001petsc}%
  \BibitemOpen
  \bibfield  {author} {\bibinfo {author} {\bibnamefont {Balay}, \bibfnamefont
  {S.}}, \bibinfo {author} {\bibnamefont {Abhyankar}, \bibfnamefont {S.}},
  \bibinfo {author} {\bibnamefont {Adams}, \bibfnamefont {M.}}, \bibinfo
  {author} {\bibnamefont {Benson}, \bibfnamefont {S.}}, \bibinfo {author}
  {\bibnamefont {Brown}, \bibfnamefont {J.}}, \bibinfo {author} {\bibnamefont
  {Brune}, \bibfnamefont {P.}}, \bibinfo {author} {\bibnamefont {Buschelman},
  \bibfnamefont {K.}}, \bibinfo {author} {\bibnamefont {Constantinescu},
  \bibfnamefont {E.}}, \bibinfo {author} {\bibnamefont {Dalcin}, \bibfnamefont
  {L.}}, \bibinfo {author} {\bibnamefont {Dener}, \bibfnamefont {A.}}, \bibinfo
  {author} {\bibnamefont {Eijkhout}, \bibfnamefont {V.}}, \bibinfo {author}
  {\bibnamefont {Faibussowitsch}, \bibfnamefont {J.}}, \bibinfo {author}
  {\bibnamefont {Gropp}, \bibfnamefont {W.}}, \bibinfo {author} {\bibnamefont
  {Hapla}, \bibfnamefont {V.}}, \bibinfo {author} {\bibnamefont {Isaac},
  \bibfnamefont {T.}}, \bibinfo {author} {\bibnamefont {Jolivet}, \bibfnamefont
  {P.}}, \bibinfo {author} {\bibnamefont {Karpeev}, \bibfnamefont {D.}},
  \bibinfo {author} {\bibnamefont {Kaushik}, \bibfnamefont {D.}}, \bibinfo
  {author} {\bibnamefont {Knepley}, \bibfnamefont {M.}}, \bibinfo {author}
  {\bibnamefont {Kong}, \bibfnamefont {F.}}, \bibinfo {author} {\bibnamefont
  {Kruger}, \bibfnamefont {S.}}, \bibinfo {author} {\bibnamefont {May},
  \bibfnamefont {D.}}, \bibinfo {author} {\bibnamefont {McInnes}, \bibfnamefont
  {L.~C.}}, \bibinfo {author} {\bibnamefont {Mills}, \bibfnamefont {R.~T.}},
  \bibinfo {author} {\bibnamefont {Mitchell}, \bibfnamefont {L.}}, \bibinfo
  {author} {\bibnamefont {Munson}, \bibfnamefont {T.}}, \bibinfo {author}
  {\bibnamefont {Roman}, \bibfnamefont {J.}}, \bibinfo {author} {\bibnamefont
  {Rupp}, \bibfnamefont {K.}}, \bibinfo {author} {\bibnamefont {Sanan},
  \bibfnamefont {P.}}, \bibinfo {author} {\bibnamefont {Sarich}, \bibfnamefont
  {J.}}, \bibinfo {author} {\bibnamefont {Smith}, \bibfnamefont {B.}}, \bibinfo
  {author} {\bibnamefont {Zampini}, \bibfnamefont {S.}}, \bibinfo {author}
  {\bibnamefont {Zhang}, \bibfnamefont {H.}}, \bibinfo {author} {\bibnamefont
  {Zhang}, \bibfnamefont {H.}}, and\ \bibinfo {author} {\bibnamefont {Zhang},
  \bibfnamefont {J.}},\ }\href {https://petsc.org/} {\enquote {\bibinfo {title}
  {{{PETS}c {W}eb page}},}\ }\bibinfo {howpublished} {\url{https://petsc.org/}}
  (\bibinfo {year} {2023}{\natexlab{a}})\BibitemShut {NoStop}%
\bibitem [{\citenamefont {Balay}\ \emph
  {et~al.}(2023{\natexlab{b}})\citenamefont {Balay}, \citenamefont {Abhyankar},
  \citenamefont {Adams}, \citenamefont {Benson}, \citenamefont {Brown},
  \citenamefont {Brune}, \citenamefont {Buschelman}, \citenamefont
  {Constantinescu}, \citenamefont {Dalcin}, \citenamefont {Dener},
  \citenamefont {Eijkhout}, \citenamefont {Faibussowitsch}, \citenamefont
  {Gropp}, \citenamefont {Hapla}, \citenamefont {Isaac}, \citenamefont
  {Jolivet}, \citenamefont {Karpeev}, \citenamefont {Kaushik}, \citenamefont
  {Knepley}, \citenamefont {Kong}, \citenamefont {Kruger}, \citenamefont {May},
  \citenamefont {McInnes}, \citenamefont {Mills}, \citenamefont {Mitchell},
  \citenamefont {Munson}, \citenamefont {Roman}, \citenamefont {Rupp},
  \citenamefont {Sanan}, \citenamefont {Sarich}, \citenamefont {Smith},
  \citenamefont {Zampini}, \citenamefont {Zhang}, \citenamefont {Zhang},\ and\
  \citenamefont {Zhang}}]{Balay2010}%
  \BibitemOpen
  \bibfield  {author} {\bibinfo {author} {\bibnamefont {Balay}, \bibfnamefont
  {S.}}, \bibinfo {author} {\bibnamefont {Abhyankar}, \bibfnamefont {S.}},
  \bibinfo {author} {\bibnamefont {Adams}, \bibfnamefont {M.}}, \bibinfo
  {author} {\bibnamefont {Benson}, \bibfnamefont {S.}}, \bibinfo {author}
  {\bibnamefont {Brown}, \bibfnamefont {J.}}, \bibinfo {author} {\bibnamefont
  {Brune}, \bibfnamefont {P.}}, \bibinfo {author} {\bibnamefont {Buschelman},
  \bibfnamefont {K.}}, \bibinfo {author} {\bibnamefont {Constantinescu},
  \bibfnamefont {E.}}, \bibinfo {author} {\bibnamefont {Dalcin}, \bibfnamefont
  {L.}}, \bibinfo {author} {\bibnamefont {Dener}, \bibfnamefont {A.}}, \bibinfo
  {author} {\bibnamefont {Eijkhout}, \bibfnamefont {V.}}, \bibinfo {author}
  {\bibnamefont {Faibussowitsch}, \bibfnamefont {J.}}, \bibinfo {author}
  {\bibnamefont {Gropp}, \bibfnamefont {W.}}, \bibinfo {author} {\bibnamefont
  {Hapla}, \bibfnamefont {V.}}, \bibinfo {author} {\bibnamefont {Isaac},
  \bibfnamefont {T.}}, \bibinfo {author} {\bibnamefont {Jolivet}, \bibfnamefont
  {P.}}, \bibinfo {author} {\bibnamefont {Karpeev}, \bibfnamefont {D.}},
  \bibinfo {author} {\bibnamefont {Kaushik}, \bibfnamefont {D.}}, \bibinfo
  {author} {\bibnamefont {Knepley}, \bibfnamefont {M.}}, \bibinfo {author}
  {\bibnamefont {Kong}, \bibfnamefont {F.}}, \bibinfo {author} {\bibnamefont
  {Kruger}, \bibfnamefont {S.}}, \bibinfo {author} {\bibnamefont {May},
  \bibfnamefont {D.}}, \bibinfo {author} {\bibnamefont {McInnes}, \bibfnamefont
  {L.~C.}}, \bibinfo {author} {\bibnamefont {Mills}, \bibfnamefont {R.~T.}},
  \bibinfo {author} {\bibnamefont {Mitchell}, \bibfnamefont {L.}}, \bibinfo
  {author} {\bibnamefont {Munson}, \bibfnamefont {T.}}, \bibinfo {author}
  {\bibnamefont {Roman}, \bibfnamefont {J.}}, \bibinfo {author} {\bibnamefont
  {Rupp}, \bibfnamefont {K.}}, \bibinfo {author} {\bibnamefont {Sanan},
  \bibfnamefont {P.}}, \bibinfo {author} {\bibnamefont {Sarich}, \bibfnamefont
  {J.}}, \bibinfo {author} {\bibnamefont {Smith}, \bibfnamefont {B.}}, \bibinfo
  {author} {\bibnamefont {Zampini}, \bibfnamefont {S.}}, \bibinfo {author}
  {\bibnamefont {Zhang}, \bibfnamefont {H.}}, \bibinfo {author} {\bibnamefont
  {Zhang}, \bibfnamefont {H.}}, and\ \bibinfo {author} {\bibnamefont {Zhang},
  \bibfnamefont {J.}},\ }\href {https://doi.org/10.2172/1968587} {\enquote
  {\bibinfo {title} {{{PETSc/TAO} Users Manual}},}\ }\bibinfo {type} {Tech.
  Rep.}\ \bibinfo {number} {ANL-21/39 - Revision 3.19}\ (\bibinfo
  {institution} {Argonne National Laboratory},\ \bibinfo {year}
  {2023})\BibitemShut {NoStop}%
\bibitem [{\citenamefont {Balay}\ \emph {et~al.}(1997)\citenamefont {Balay},
  \citenamefont {Gropp}, \citenamefont {McInnes},\ and\ \citenamefont
  {Smith}}]{Balay1997}%
  \BibitemOpen
  \bibfield  {author} {\bibinfo {author} {\bibnamefont {Balay}, \bibfnamefont
  {S.}}, \bibinfo {author} {\bibnamefont {Gropp}, \bibfnamefont {W.~D.}},
  \bibinfo {author} {\bibnamefont {McInnes}, \bibfnamefont {L.~C.}}, and\
  \bibinfo {author} {\bibnamefont {Smith}, \bibfnamefont {B.~F.}},\ }\bibfield
  {title} {\enquote {\bibinfo {title} {{Efficient Management of Parallelism in
  Object-Oriented Numerical Software Libraries}},}\ }in\ \href
  {https://doi.org/10.1007/978-1-4612-1986-6_8} {\emph {\bibinfo {booktitle}
  {Modern Software Tools for Scientific Computing}}}\ (\bibinfo  {publisher}
  {Birkh{\"{a}}user, Boston, MA},\ \bibinfo {year} {1997})\BibitemShut
  {NoStop}%
\bibitem [{\citenamefont {Bhalla}\ \emph {et~al.}(2013)\citenamefont {Bhalla},
  \citenamefont {Bale}, \citenamefont {Griffith},\ and\ \citenamefont
  {Patankar}}]{Bhalla2013}%
  \BibitemOpen
  \bibfield  {author} {\bibinfo {author} {\bibnamefont {Bhalla}, \bibfnamefont
  {A.~P.~S.}}, \bibinfo {author} {\bibnamefont {Bale}, \bibfnamefont {R.}},
  \bibinfo {author} {\bibnamefont {Griffith}, \bibfnamefont {B.~E.}}, and\
  \bibinfo {author} {\bibnamefont {Patankar}, \bibfnamefont {N.~A.}},\
  }\bibfield  {title} {\enquote {\bibinfo {title} {{A unified mathematical
  framework and an adaptive numerical method for fluid-structure interaction
  with rigid, deforming, and elastic bodies}},}\ }\href
  {https://doi.org/10.1016/j.jcp.2013.04.033} {\bibfield  {journal} {\bibinfo
  {journal} {Journal of Computational Physics}\ }\textbf {\bibinfo {volume}
  {250}} (\bibinfo {year} {2013}),\ 10.1016/j.jcp.2013.04.033}\BibitemShut
  {NoStop}%
\bibitem [{\citenamefont {Bhalla}\ \emph {et~al.}(2014)\citenamefont {Bhalla},
  \citenamefont {Bale}, \citenamefont {Griffith},\ and\ \citenamefont
  {Patankar}}]{bhalla2014fully}%
  \BibitemOpen
  \bibfield  {author} {\bibinfo {author} {\bibnamefont {Bhalla}, \bibfnamefont
  {A.~P.~S.}}, \bibinfo {author} {\bibnamefont {Bale}, \bibfnamefont {R.}},
  \bibinfo {author} {\bibnamefont {Griffith}, \bibfnamefont {B.~E.}}, and\
  \bibinfo {author} {\bibnamefont {Patankar}, \bibfnamefont {N.~A.}},\
  }\bibfield  {title} {\enquote {\bibinfo {title} {{Fully resolved immersed
  electrohydrodynamics for particle motion, electrolocation, and
  self-propulsion}},}\ }\href@noop {} {\bibfield  {journal} {\bibinfo
  {journal} {Journal of Computational Physics}\ }\textbf {\bibinfo {volume}
  {256}},\ \bibinfo {pages} {88--108} (\bibinfo {year} {2014})}\BibitemShut
  {NoStop}%
\bibitem [{\citenamefont {Bhalla}, \citenamefont {Griffith},\ and\
  \citenamefont {Patankar}(2013)}]{Bhalla2013a}%
  \BibitemOpen
  \bibfield  {author} {\bibinfo {author} {\bibnamefont {Bhalla}, \bibfnamefont
  {A.~P.~S.}}, \bibinfo {author} {\bibnamefont {Griffith}, \bibfnamefont
  {B.~E.}}, and\ \bibinfo {author} {\bibnamefont {Patankar}, \bibfnamefont
  {N.~A.}},\ }\bibfield  {title} {\enquote {\bibinfo {title} {{A Forced Damped
  Oscillation Framework for Undulatory Swimming Provides New Insights into How
  Propulsion Arises in Active and Passive Swimming}},}\ }\href
  {https://doi.org/10.1371/journal.pcbi.1003097} {\bibfield  {journal}
  {\bibinfo  {journal} {PLoS Computational Biology}\ }\textbf {\bibinfo
  {volume} {9}} (\bibinfo {year} {2013}),\
  10.1371/journal.pcbi.1003097}\BibitemShut {NoStop}%
\bibitem [{\citenamefont {Bhalla}\ \emph {et~al.}(2020)\citenamefont {Bhalla},
  \citenamefont {Nangia}, \citenamefont {Dafnakis}, \citenamefont {Bracco},\
  and\ \citenamefont {Mattiazzo}}]{Bhalla2020}%
  \BibitemOpen
  \bibfield  {author} {\bibinfo {author} {\bibnamefont {Bhalla}, \bibfnamefont
  {A.~P.~S.}}, \bibinfo {author} {\bibnamefont {Nangia}, \bibfnamefont {N.}},
  \bibinfo {author} {\bibnamefont {Dafnakis}, \bibfnamefont {P.}}, \bibinfo
  {author} {\bibnamefont {Bracco}, \bibfnamefont {G.}}, and\ \bibinfo {author}
  {\bibnamefont {Mattiazzo}, \bibfnamefont {G.}},\ }\bibfield  {title}
  {\enquote {\bibinfo {title} {{Simulating water-entry/exit problems using
  Eulerian–Lagrangian and fully-Eulerian fictitious domain methods within the
  open-source IBAMR library}},}\ }\href
  {https://doi.org/10.1016/j.apor.2019.101932} {\bibfield  {journal} {\bibinfo
  {journal} {Applied Ocean Research}\ }\textbf {\bibinfo {volume} {94}}
  (\bibinfo {year} {2020}),\ 10.1016/j.apor.2019.101932}\BibitemShut {NoStop}%
\bibitem [{\citenamefont {Borazjani}\ and\ \citenamefont
  {Sotiropoulos}(2008)}]{Borazjani2008}%
  \BibitemOpen
  \bibfield  {author} {\bibinfo {author} {\bibnamefont {Borazjani},
  \bibfnamefont {I.}}and\ \bibinfo {author} {\bibnamefont {Sotiropoulos},
  \bibfnamefont {F.}},\ }\bibfield  {title} {\enquote {\bibinfo {title}
  {{Numerical investigation of the hydrodynamics of carangiform swimming in the
  transitional and inertial flow regimes}},}\ }\href
  {https://doi.org/10.1242/jeb.015644} {\bibfield  {journal} {\bibinfo
  {journal} {Journal of Experimental Biology}\ }\textbf {\bibinfo {volume}
  {211}} (\bibinfo {year} {2008}),\ 10.1242/jeb.015644}\BibitemShut {NoStop}%
\bibitem [{\citenamefont {Borazjani}\ and\ \citenamefont
  {Sotiropoulos}(2009)}]{Borazjani2009}%
  \BibitemOpen
  \bibfield  {author} {\bibinfo {author} {\bibnamefont {Borazjani},
  \bibfnamefont {I.}}and\ \bibinfo {author} {\bibnamefont {Sotiropoulos},
  \bibfnamefont {F.}},\ }\bibfield  {title} {\enquote {\bibinfo {title}
  {{Numerical investigation of the hydrodynamics of anguilliform swimming in
  the transitional and inertial flow regimes}},}\ }\href
  {https://doi.org/10.1242/jeb.025007} {\bibfield  {journal} {\bibinfo
  {journal} {Journal of Experimental Biology}\ }\textbf {\bibinfo {volume}
  {212}} (\bibinfo {year} {2009}),\ 10.1242/jeb.025007}\BibitemShut {NoStop}%
\bibitem [{\citenamefont {Borazjani}\ and\ \citenamefont
  {Sotiropoulos}(2010)}]{Borazjani2010}%
  \BibitemOpen
  \bibfield  {author} {\bibinfo {author} {\bibnamefont {Borazjani},
  \bibfnamefont {I.}}and\ \bibinfo {author} {\bibnamefont {Sotiropoulos},
  \bibfnamefont {F.}},\ }\bibfield  {title} {\enquote {\bibinfo {title} {{On
  the role of form and kinematics on the hydrodynamics of self-propelled
  body/caudal fin swimming}},}\ }\href {https://doi.org/10.1242/jeb.030932}
  {\bibfield  {journal} {\bibinfo  {journal} {Journal of Experimental Biology}\
  }\textbf {\bibinfo {volume} {213}} (\bibinfo {year} {2010}),\
  10.1242/jeb.030932}\BibitemShut {NoStop}%
\bibitem [{\citenamefont {Carling}, \citenamefont {Williams},\ and\
  \citenamefont {Bowtell}(1998)}]{Carling1998}%
  \BibitemOpen
  \bibfield  {author} {\bibinfo {author} {\bibnamefont {Carling}, \bibfnamefont
  {J.}}, \bibinfo {author} {\bibnamefont {Williams}, \bibfnamefont {T.~L.}},
  and\ \bibinfo {author} {\bibnamefont {Bowtell}, \bibfnamefont {G.}},\
  }\bibfield  {title} {\enquote {\bibinfo {title} {{Self-propelled anguilliform
  swimming: Simultaneous solution of the two-dimensional Navier-Stokes
  equations and Newton's laws of motion}},}\ }\href
  {https://doi.org/10.1242/jeb.201.23.3143} {\bibfield  {journal} {\bibinfo
  {journal} {Journal of Experimental Biology}\ }\textbf {\bibinfo {volume}
  {201}} (\bibinfo {year} {1998}),\ 10.1242/jeb.201.23.3143}\BibitemShut
  {NoStop}%
\bibitem [{\citenamefont {Chao}, \citenamefont {Alam},\ and\ \citenamefont
  {Cheng}(2022)}]{chao2022hydrodynamic}%
  \BibitemOpen
  \bibfield  {author} {\bibinfo {author} {\bibnamefont {Chao}, \bibfnamefont
  {L.-M.}}, \bibinfo {author} {\bibnamefont {Alam}, \bibfnamefont {M.~M.}},
  and\ \bibinfo {author} {\bibnamefont {Cheng}, \bibfnamefont {L.}},\
  }\bibfield  {title} {\enquote {\bibinfo {title} {{Hydrodynamic performance of
  slender swimmer: effect of travelling wavelength}},}\ }\href@noop {}
  {\bibfield  {journal} {\bibinfo  {journal} {Journal of Fluid Mechanics}\
  }\textbf {\bibinfo {volume} {947}},\ \bibinfo {pages} {A8} (\bibinfo {year}
  {2022})}\BibitemShut {NoStop}%
\bibitem [{\citenamefont {Chao}, \citenamefont {Alam},\ and\ \citenamefont
  {Ji}(2021)}]{Chao2021}%
  \BibitemOpen
  \bibfield  {author} {\bibinfo {author} {\bibnamefont {Chao}, \bibfnamefont
  {L.~M.}}, \bibinfo {author} {\bibnamefont {Alam}, \bibfnamefont {M.~M.}},
  and\ \bibinfo {author} {\bibnamefont {Ji}, \bibfnamefont {C.}},\ }\bibfield
  {title} {\enquote {\bibinfo {title} {{Drag-thrust transition and wake
  structures of a pitching foil undergoing asymmetric oscillation}},}\ }\href
  {https://doi.org/10.1016/j.jfluidstructs.2021.103289} {\bibfield  {journal}
  {\bibinfo  {journal} {Journal of Fluids and Structures}\ }\textbf {\bibinfo
  {volume} {103}} (\bibinfo {year} {2021}),\
  10.1016/j.jfluidstructs.2021.103289}\BibitemShut {NoStop}%
\bibitem [{\citenamefont {Chao}\ \emph {et~al.}(2019)\citenamefont {Chao},
  \citenamefont {Pan}, \citenamefont {Zhang},\ and\ \citenamefont
  {Yan}}]{Chao2019}%
  \BibitemOpen
  \bibfield  {author} {\bibinfo {author} {\bibnamefont {Chao}, \bibfnamefont
  {L.~M.}}, \bibinfo {author} {\bibnamefont {Pan}, \bibfnamefont {G.}},
  \bibinfo {author} {\bibnamefont {Zhang}, \bibfnamefont {D.}}, and\ \bibinfo
  {author} {\bibnamefont {Yan}, \bibfnamefont {G.~X.}},\ }\bibfield  {title}
  {\enquote {\bibinfo {title} {{On the two staggered swimming fish}},}\ }\href
  {https://doi.org/10.1016/j.chaos.2019.04.028} {\bibfield  {journal} {\bibinfo
   {journal} {Chaos, Solitons and Fractals}\ }\textbf {\bibinfo {volume} {123}}
  (\bibinfo {year} {2019}),\ 10.1016/j.chaos.2019.04.028}\BibitemShut {NoStop}%
\bibitem [{\citenamefont {Coombs}\ and\ \citenamefont
  {Montgomery}(2014)}]{Coombs2014}%
  \BibitemOpen
  \bibfield  {author} {\bibinfo {author} {\bibnamefont {Coombs}, \bibfnamefont
  {S.}}and\ \bibinfo {author} {\bibnamefont {Montgomery}, \bibfnamefont {J.}},\
  }\bibfield  {title} {\enquote {\bibinfo {title} {{The role of flow and the
  lateral line in the multisensory guidance of orienting behaviors}},}\ }in\
  \href {https://doi.org/10.1007/978-3-642-41446-6_3} {\emph {\bibinfo
  {booktitle} {Flow Sensing in Air and Water: Behavioral, Neural and
  Engineering Principles of Operation}}}\ (\bibinfo  {publisher} {Springer,
  Berlin, Heidelberg},\ \bibinfo {year} {2014})\BibitemShut {NoStop}%
\bibitem [{\citenamefont {Cranmer}(2020)}]{pysr}%
  \BibitemOpen
  \bibfield  {author} {\bibinfo {author} {\bibnamefont {Cranmer}, \bibfnamefont
  {M.}},\ }\href {https://doi.org/10.5281/zenodo.4041459} {\enquote {\bibinfo
  {title} {{PySR: Fast \& Parallelized Symbolic Regression in Python/Julia}},}\
  } (\bibinfo {year} {2020})\BibitemShut {NoStop}%
\bibitem [{\citenamefont {Deng}\ and\ \citenamefont {Liu}(2021)}]{Deng2021}%
  \BibitemOpen
  \bibfield  {author} {\bibinfo {author} {\bibnamefont {Deng}, \bibfnamefont
  {J.}}and\ \bibinfo {author} {\bibnamefont {Liu}, \bibfnamefont {D.}},\
  }\bibfield  {title} {\enquote {\bibinfo {title} {{Spontaneous response of a
  self-organized fish school to a predator}},}\ }\href
  {https://doi.org/10.1088/1748-3190/abfd7f} {\bibfield  {journal} {\bibinfo
  {journal} {Bioinspiration and Biomimetics}\ }\textbf {\bibinfo {volume} {16}}
  (\bibinfo {year} {2021}),\ 10.1088/1748-3190/abfd7f}\BibitemShut {NoStop}%
\bibitem [{\citenamefont {Deng}, \citenamefont {Shao},\ and\ \citenamefont
  {Yu}(2007)}]{Deng2007}%
  \BibitemOpen
  \bibfield  {author} {\bibinfo {author} {\bibnamefont {Deng}, \bibfnamefont
  {J.}}, \bibinfo {author} {\bibnamefont {Shao}, \bibfnamefont {X.~M.}}, and\
  \bibinfo {author} {\bibnamefont {Yu}, \bibfnamefont {Z.~S.}},\ }\bibfield
  {title} {\enquote {\bibinfo {title} {{Hydrodynamic studies on two traveling
  wavy foils in tandem arrangement}},}\ }\href
  {https://doi.org/10.1063/1.2814259} {\bibfield  {journal} {\bibinfo
  {journal} {Physics of Fluids}\ }\textbf {\bibinfo {volume} {19}} (\bibinfo
  {year} {2007}),\ 10.1063/1.2814259}\BibitemShut {NoStop}%
\bibitem [{\citenamefont {Deng}\ \emph {et~al.}(2016)\citenamefont {Deng},
  \citenamefont {Sun}, \citenamefont {{Lubao Teng}}, \citenamefont {Pan},\ and\
  \citenamefont {Shao}}]{Deng2016}%
  \BibitemOpen
  \bibfield  {author} {\bibinfo {author} {\bibnamefont {Deng}, \bibfnamefont
  {J.}}, \bibinfo {author} {\bibnamefont {Sun}, \bibfnamefont {L.}}, \bibinfo
  {author} {\bibnamefont {{Lubao Teng}},}, \bibinfo {author} {\bibnamefont
  {Pan}, \bibfnamefont {D.}}, and\ \bibinfo {author} {\bibnamefont {Shao},
  \bibfnamefont {X.}},\ }\bibfield  {title} {\enquote {\bibinfo {title} {{The
  correlation between wake transition and propulsive efficiency of a flapping
  foil: A numerical study}},}\ }\href {https://doi.org/10.1063/1.4961566}
  {\bibfield  {journal} {\bibinfo  {journal} {Physics of Fluids}\ }\textbf
  {\bibinfo {volume} {28}} (\bibinfo {year} {2016}),\
  10.1063/1.4961566}\BibitemShut {NoStop}%
\bibitem [{\citenamefont {Deng}\ \emph {et~al.}(2015)\citenamefont {Deng},
  \citenamefont {Teng}, \citenamefont {Pan},\ and\ \citenamefont
  {Shao}}]{Deng2015}%
  \BibitemOpen
  \bibfield  {author} {\bibinfo {author} {\bibnamefont {Deng}, \bibfnamefont
  {J.}}, \bibinfo {author} {\bibnamefont {Teng}, \bibfnamefont {L.}}, \bibinfo
  {author} {\bibnamefont {Pan}, \bibfnamefont {D.}}, and\ \bibinfo {author}
  {\bibnamefont {Shao}, \bibfnamefont {X.}},\ }\bibfield  {title} {\enquote
  {\bibinfo {title} {{Inertial effects of the semi-passive flapping foil on its
  energy extraction efficiency}},}\ }\href {https://doi.org/10.1063/1.4921384}
  {\bibfield  {journal} {\bibinfo  {journal} {Physics of Fluids}\ }\textbf
  {\bibinfo {volume} {27}} (\bibinfo {year} {2015}),\
  10.1063/1.4921384}\BibitemShut {NoStop}%
\bibitem [{\citenamefont {Deng}\ \emph {et~al.}(2022)\citenamefont {Deng},
  \citenamefont {Wang}, \citenamefont {Kandel},\ and\ \citenamefont
  {Teng}}]{Deng2022}%
  \BibitemOpen
  \bibfield  {author} {\bibinfo {author} {\bibnamefont {Deng}, \bibfnamefont
  {J.}}, \bibinfo {author} {\bibnamefont {Wang}, \bibfnamefont {S.}}, \bibinfo
  {author} {\bibnamefont {Kandel}, \bibfnamefont {P.}}, and\ \bibinfo {author}
  {\bibnamefont {Teng}, \bibfnamefont {L.}},\ }\bibfield  {title} {\enquote
  {\bibinfo {title} {{Effects of free surface on a flapping-foil based ocean
  current energy extractor}},}\ }\href
  {https://doi.org/10.1016/j.renene.2021.09.098} {\bibfield  {journal}
  {\bibinfo  {journal} {Renewable Energy}\ }\textbf {\bibinfo {volume} {181}}
  (\bibinfo {year} {2022}),\ 10.1016/j.renene.2021.09.098}\BibitemShut
  {NoStop}%
\bibitem [{\citenamefont {Dewey}\ \emph {et~al.}(2014)\citenamefont {Dewey},
  \citenamefont {Quinn}, \citenamefont {Boschitsch},\ and\ \citenamefont
  {Smits}}]{Dewey2014a}%
  \BibitemOpen
  \bibfield  {author} {\bibinfo {author} {\bibnamefont {Dewey}, \bibfnamefont
  {P.~A.}}, \bibinfo {author} {\bibnamefont {Quinn}, \bibfnamefont {D.~B.}},
  \bibinfo {author} {\bibnamefont {Boschitsch}, \bibfnamefont {B.~M.}}, and\
  \bibinfo {author} {\bibnamefont {Smits}, \bibfnamefont {A.~J.}},\ }\bibfield
  {title} {\enquote {\bibinfo {title} {{Propulsive performance of unsteady
  tandem hydrofoils in a side-by-side configuration}},}\ }\href
  {https://doi.org/10.1063/1.4871024} {\bibfield  {journal} {\bibinfo
  {journal} {Physics of Fluids}\ }\textbf {\bibinfo {volume} {26}} (\bibinfo
  {year} {2014}),\ 10.1063/1.4871024}\BibitemShut {NoStop}%
\bibitem [{\citenamefont {{Du Clos}}\ \emph {et~al.}(2019)\citenamefont {{Du
  Clos}}, \citenamefont {Dabiri}, \citenamefont {Costello}, \citenamefont
  {Colin}, \citenamefont {Morgan}, \citenamefont {Fogerson},\ and\
  \citenamefont {Gemmell}}]{DuClos2019}%
  \BibitemOpen
  \bibfield  {author} {\bibinfo {author} {\bibnamefont {{Du Clos}},
  \bibfnamefont {K.~T.}}, \bibinfo {author} {\bibnamefont {Dabiri},
  \bibfnamefont {J.~O.}}, \bibinfo {author} {\bibnamefont {Costello},
  \bibfnamefont {J.~H.}}, \bibinfo {author} {\bibnamefont {Colin},
  \bibfnamefont {S.~P.}}, \bibinfo {author} {\bibnamefont {Morgan},
  \bibfnamefont {J.~R.}}, \bibinfo {author} {\bibnamefont {Fogerson},
  \bibfnamefont {S.~M.}}, and\ \bibinfo {author} {\bibnamefont {Gemmell},
  \bibfnamefont {B.~J.}},\ }\bibfield  {title} {\enquote {\bibinfo {title}
  {{Thrust generation during steady swimming and acceleration from rest in
  anguilliform swimmers}},}\ }\href {https://doi.org/10.1242/jeb.212464}
  {\bibfield  {journal} {\bibinfo  {journal} {Journal of Experimental Biology}\
  }\textbf {\bibinfo {volume} {222}} (\bibinfo {year} {2019}),\
  10.1242/jeb.212464}\BibitemShut {NoStop}%
\bibitem [{\citenamefont {Falgout}\ \emph {et~al.}(2010)\citenamefont
  {Falgout}, \citenamefont {Cleary}, \citenamefont {Jones}, \citenamefont
  {Chow}, \citenamefont {Henson}, \citenamefont {Baldwin}, \citenamefont
  {Brown}, \citenamefont {Vassilevski},\ and\ \citenamefont
  {Yang}}]{falgout2010hypre}%
  \BibitemOpen
  \bibfield  {author} {\bibinfo {author} {\bibnamefont {Falgout}, \bibfnamefont
  {R.}}, \bibinfo {author} {\bibnamefont {Cleary}, \bibfnamefont {A.}},
  \bibinfo {author} {\bibnamefont {Jones}, \bibfnamefont {J.}}, \bibinfo
  {author} {\bibnamefont {Chow}, \bibfnamefont {E.}}, \bibinfo {author}
  {\bibnamefont {Henson}, \bibfnamefont {V.}}, \bibinfo {author} {\bibnamefont
  {Baldwin}, \bibfnamefont {C.}}, \bibinfo {author} {\bibnamefont {Brown},
  \bibfnamefont {P.}}, \bibinfo {author} {\bibnamefont {Vassilevski},
  \bibfnamefont {P.}}, and\ \bibinfo {author} {\bibnamefont {Yang},
  \bibfnamefont {U.~M.}},\ }\bibfield  {title} {\enquote {\bibinfo {title}
  {{HYPRE: High Performance Preconditioners}},}\ }\href@noop {} {\bibfield
  {journal} {\bibinfo  {journal} {Users Manual. Version}\ }\textbf {\bibinfo
  {volume} {1}} (\bibinfo {year} {2010})}\BibitemShut {NoStop}%
\bibitem [{\citenamefont {Fish}(2020)}]{Fish2020}%
  \BibitemOpen
  \bibfield  {author} {\bibinfo {author} {\bibnamefont {Fish}, \bibfnamefont
  {F.~E.}},\ }\bibfield  {title} {\enquote {\bibinfo {title} {{Advantages of
  aquatic animals as models for bio-inspired drones over present AUV
  technology}},}\ }\href {https://doi.org/10.1088/1748-3190/ab5a34} {\bibfield
  {journal} {\bibinfo  {journal} {Bioinspiration and Biomimetics}\ }\textbf
  {\bibinfo {volume} {15}} (\bibinfo {year} {2020}),\
  10.1088/1748-3190/ab5a34}\BibitemShut {NoStop}%
\bibitem [{\citenamefont {Gazzola}, \citenamefont {Argentina},\ and\
  \citenamefont {Mahadevan}(2014)}]{Gazzola2014}%
  \BibitemOpen
  \bibfield  {author} {\bibinfo {author} {\bibnamefont {Gazzola}, \bibfnamefont
  {M.}}, \bibinfo {author} {\bibnamefont {Argentina}, \bibfnamefont {M.}}, and\
  \bibinfo {author} {\bibnamefont {Mahadevan}, \bibfnamefont {L.}},\ }\bibfield
   {title} {\enquote {\bibinfo {title} {{Scaling macroscopic aquatic
  locomotion}},}\ }\href {https://doi.org/10.1038/nphys3078} {\bibfield
  {journal} {\bibinfo  {journal} {Nature Physics}\ }\textbf {\bibinfo {volume}
  {10}} (\bibinfo {year} {2014}),\ 10.1038/nphys3078}\BibitemShut {NoStop}%
\bibitem [{\citenamefont {Gazzola}\ \emph {et~al.}(2012)\citenamefont
  {Gazzola}, \citenamefont {Mimeau}, \citenamefont {Tchieu},\ and\
  \citenamefont {Koumoutsakos}}]{Gazzola2012}%
  \BibitemOpen
  \bibfield  {author} {\bibinfo {author} {\bibnamefont {Gazzola}, \bibfnamefont
  {M.}}, \bibinfo {author} {\bibnamefont {Mimeau}, \bibfnamefont {C.}},
  \bibinfo {author} {\bibnamefont {Tchieu}, \bibfnamefont {A.~A.}}, and\
  \bibinfo {author} {\bibnamefont {Koumoutsakos}, \bibfnamefont {P.}},\
  }\bibfield  {title} {\enquote {\bibinfo {title} {{Flow Mediated Interactions
  Between Two Cylinders at Finite $Re$ Numbers}},}\ }\href
  {https://doi.org/10.1063/1.4704195} {\bibfield  {journal} {\bibinfo
  {journal} {Physics of Fluids}\ }\textbf {\bibinfo {volume} {24}},\ \bibinfo
  {pages} {043103} (\bibinfo {year} {2012})}\BibitemShut {NoStop}%
\bibitem [{\citenamefont {Griffith}(2013)}]{griffith2013ibamr}%
  \BibitemOpen
  \bibfield  {author} {\bibinfo {author} {\bibnamefont {Griffith},
  \bibfnamefont {B.~E.}},\ }\bibfield  {title} {\enquote {\bibinfo {title}
  {{IBAMR: An adaptive and distributed-memory parallel implementation of the
  immersed boundary method}},}\ }\href@noop {} {\bibfield  {journal} {\bibinfo
  {journal} {URL https://ibamr. github. io/about}\ } (\bibinfo {year}
  {2013})}\BibitemShut {NoStop}%
\bibitem [{\citenamefont {Griffith}\ and\ \citenamefont
  {Patankar}(2020)}]{Griffith2020}%
  \BibitemOpen
  \bibfield  {author} {\bibinfo {author} {\bibnamefont {Griffith},
  \bibfnamefont {B.~E.}}and\ \bibinfo {author} {\bibnamefont {Patankar},
  \bibfnamefont {N.~A.}},\ }\href
  {https://doi.org/10.1146/annurev-fluid-010719-060228} {\enquote {\bibinfo
  {title} {{Immersed Methods for Fluid-Structure Interaction}},}\ } (\bibinfo
  {year} {2020})\BibitemShut {NoStop}%
\bibitem [{\citenamefont {Gungor}\ and\ \citenamefont
  {Hemmati}(2021)}]{Gungor2021}%
  \BibitemOpen
  \bibfield  {author} {\bibinfo {author} {\bibnamefont {Gungor}, \bibfnamefont
  {A.}}and\ \bibinfo {author} {\bibnamefont {Hemmati}, \bibfnamefont {A.}},\
  }\bibfield  {title} {\enquote {\bibinfo {title} {{The scaling and performance
  of side-by-side pitching hydrofoils}},}\ }\href
  {https://doi.org/10.1016/j.jfluidstructs.2021.103320} {\bibfield  {journal}
  {\bibinfo  {journal} {Journal of Fluids and Structures}\ }\textbf {\bibinfo
  {volume} {104}} (\bibinfo {year} {2021}),\
  10.1016/j.jfluidstructs.2021.103320}\BibitemShut {NoStop}%
\bibitem [{\citenamefont {Gupta}\ \emph {et~al.}(2021)\citenamefont {Gupta},
  \citenamefont {Thekkethil}, \citenamefont {Agrawal}, \citenamefont
  {Hourigan}, \citenamefont {Thompson},\ and\ \citenamefont
  {Sharma}}]{Gupta2021}%
  \BibitemOpen
  \bibfield  {author} {\bibinfo {author} {\bibnamefont {Gupta}, \bibfnamefont
  {S.}}, \bibinfo {author} {\bibnamefont {Thekkethil}, \bibfnamefont {N.}},
  \bibinfo {author} {\bibnamefont {Agrawal}, \bibfnamefont {A.}}, \bibinfo
  {author} {\bibnamefont {Hourigan}, \bibfnamefont {K.}}, \bibinfo {author}
  {\bibnamefont {Thompson}, \bibfnamefont {M.~C.}}, and\ \bibinfo {author}
  {\bibnamefont {Sharma}, \bibfnamefont {A.}},\ }\bibfield  {title} {\enquote
  {\bibinfo {title} {{Body-caudal fin fish-inspired self-propulsion study on
  burst-and-coast and continuous swimming of a hydrofoil model}},}\ }\href
  {https://doi.org/10.1063/5.0061417} {\bibfield  {journal} {\bibinfo
  {journal} {Physics of Fluids}\ }\textbf {\bibinfo {volume} {33}} (\bibinfo
  {year} {2021}),\ 10.1063/5.0061417}\BibitemShut {NoStop}%
\bibitem [{\citenamefont {Hornung}\ and\ \citenamefont
  {Kohn}(2002)}]{Hornung2002}%
  \BibitemOpen
  \bibfield  {author} {\bibinfo {author} {\bibnamefont {Hornung}, \bibfnamefont
  {R.~D.}}and\ \bibinfo {author} {\bibnamefont {Kohn}, \bibfnamefont {S.~R.}},\
  }\bibfield  {title} {\enquote {\bibinfo {title} {{Managing application
  complexity in the SAMRAI object-oriented framework}},}\ }\href
  {https://doi.org/10.1002/cpe.652} {\bibfield  {journal} {\bibinfo  {journal}
  {Concurrency and Computation: Practice and Experience}\ }\textbf {\bibinfo
  {volume} {14}} (\bibinfo {year} {2002}),\ 10.1002/cpe.652}\BibitemShut
  {NoStop}%
\bibitem [{\citenamefont {Hornung}, \citenamefont {Wissink},\ and\
  \citenamefont {Kohn}(2006)}]{Hornung2006}%
  \BibitemOpen
  \bibfield  {author} {\bibinfo {author} {\bibnamefont {Hornung}, \bibfnamefont
  {R.~D.}}, \bibinfo {author} {\bibnamefont {Wissink}, \bibfnamefont {A.~M.}},
  and\ \bibinfo {author} {\bibnamefont {Kohn}, \bibfnamefont {S.~R.}},\
  }\bibfield  {title} {\enquote {\bibinfo {title} {{Managing complex data and
  geometry in parallel structured AMR applications}},}\ }\href
  {https://doi.org/10.1007/s00366-006-0038-6} {\bibfield  {journal} {\bibinfo
  {journal} {Engineering with Computers}\ }\textbf {\bibinfo {volume} {22}}
  (\bibinfo {year} {2006}),\ 10.1007/s00366-006-0038-6}\BibitemShut {NoStop}%
\bibitem [{\citenamefont {Huera-Huarte}(2018)}]{Huera-Huarte2018}%
  \BibitemOpen
  \bibfield  {author} {\bibinfo {author} {\bibnamefont {Huera-Huarte},
  \bibfnamefont {F.~J.}},\ }\bibfield  {title} {\enquote {\bibinfo {title}
  {{Propulsive performance of a pair of pitching foils in staggered
  configurations}},}\ }\href
  {https://doi.org/10.1016/j.jfluidstructs.2018.04.024} {\bibfield  {journal}
  {\bibinfo  {journal} {Journal of Fluids and Structures}\ }\textbf {\bibinfo
  {volume} {81}} (\bibinfo {year} {2018}),\
  10.1016/j.jfluidstructs.2018.04.024}\BibitemShut {NoStop}%
\bibitem [{\citenamefont {Khalid}\ \emph {et~al.}(2021)\citenamefont {Khalid},
  \citenamefont {Wang}, \citenamefont {Akhtar}, \citenamefont {Dong},
  \citenamefont {Liu},\ and\ \citenamefont {Hemmati}}]{Khalid2021}%
  \BibitemOpen
  \bibfield  {author} {\bibinfo {author} {\bibnamefont {Khalid}, \bibfnamefont
  {M.~S.~U.}}, \bibinfo {author} {\bibnamefont {Wang}, \bibfnamefont {J.}},
  \bibinfo {author} {\bibnamefont {Akhtar}, \bibfnamefont {I.}}, \bibinfo
  {author} {\bibnamefont {Dong}, \bibfnamefont {H.}}, \bibinfo {author}
  {\bibnamefont {Liu}, \bibfnamefont {M.}}, and\ \bibinfo {author}
  {\bibnamefont {Hemmati}, \bibfnamefont {A.}},\ }\bibfield  {title} {\enquote
  {\bibinfo {title} {{Why do anguilliform swimmers perform undulation with
  wavelengths shorter than their bodylengths?}}}\ }\href
  {https://doi.org/10.1063/5.0040473} {\bibfield  {journal} {\bibinfo
  {journal} {Physics of Fluids}\ }\textbf {\bibinfo {volume} {33}} (\bibinfo
  {year} {2021}),\ 10.1063/5.0040473}\BibitemShut {NoStop}%
\bibitem [{\citenamefont {Khalid}\ \emph {et~al.}(2020)\citenamefont {Khalid},
  \citenamefont {Wang}, \citenamefont {Dong},\ and\ \citenamefont
  {Liu}}]{Khalid2020}%
  \BibitemOpen
  \bibfield  {author} {\bibinfo {author} {\bibnamefont {Khalid}, \bibfnamefont
  {M.~S.~U.}}, \bibinfo {author} {\bibnamefont {Wang}, \bibfnamefont {J.}},
  \bibinfo {author} {\bibnamefont {Dong}, \bibfnamefont {H.}}, and\ \bibinfo
  {author} {\bibnamefont {Liu}, \bibfnamefont {M.}},\ }\bibfield  {title}
  {\enquote {\bibinfo {title} {{Flow transitions and mapping for undulating
  swimmers}},}\ }\href {https://doi.org/10.1103/PhysRevFluids.5.063104}
  {\bibfield  {journal} {\bibinfo  {journal} {Physical Review Fluids}\ }\textbf
  {\bibinfo {volume} {5}} (\bibinfo {year} {2020}),\
  10.1103/PhysRevFluids.5.063104}\BibitemShut {NoStop}%
\bibitem [{\citenamefont {Kirk}\ \emph {et~al.}(2006)\citenamefont {Kirk},
  \citenamefont {Peterson}, \citenamefont {Stogner},\ and\ \citenamefont
  {Carey}}]{Kirk2006}%
  \BibitemOpen
  \bibfield  {author} {\bibinfo {author} {\bibnamefont {Kirk}, \bibfnamefont
  {B.~S.}}, \bibinfo {author} {\bibnamefont {Peterson}, \bibfnamefont {J.~W.}},
  \bibinfo {author} {\bibnamefont {Stogner}, \bibfnamefont {R.~H.}}, and\
  \bibinfo {author} {\bibnamefont {Carey}, \bibfnamefont {G.~F.}},\ }\bibfield
  {title} {\enquote {\bibinfo {title} {{libMesh : a C++ library for parallel
  adaptive mesh refinement/coarsening simulations}},}\ }\href
  {https://doi.org/10.1007/s00366-006-0049-3} {\bibfield  {journal} {\bibinfo
  {journal} {Engineering with Computers}\ }\textbf {\bibinfo {volume} {22}}
  (\bibinfo {year} {2006}),\ 10.1007/s00366-006-0049-3}\BibitemShut {NoStop}%
\bibitem [{\citenamefont {Lamb}(1932)}]{Hlamb1932hydrodynamics}%
  \BibitemOpen
  \bibfield  {author} {\bibinfo {author} {\bibnamefont {Lamb}, \bibfnamefont
  {H.}},\ }\href@noop {} {\emph {\bibinfo {title} {{Hydrodynamics}}}},\
  \bibinfo {edition} {6th}\ ed.\ (\bibinfo  {publisher} {Cambridge University
  Press},\ \bibinfo {address} {Cambridge},\ \bibinfo {year} {1932})\ p.\
  \bibinfo {pages} {182}\BibitemShut {NoStop}%
\bibitem [{\citenamefont {Lecheval}\ \emph {et~al.}(2018)\citenamefont
  {Lecheval}, \citenamefont {Jiang}, \citenamefont {Tichit}, \citenamefont
  {Sire}, \citenamefont {Hemelrijk},\ and\ \citenamefont
  {Theraulaz}}]{Lecheval2018}%
  \BibitemOpen
  \bibfield  {author} {\bibinfo {author} {\bibnamefont {Lecheval},
  \bibfnamefont {V.}}, \bibinfo {author} {\bibnamefont {Jiang}, \bibfnamefont
  {L.}}, \bibinfo {author} {\bibnamefont {Tichit}, \bibfnamefont {P.}},
  \bibinfo {author} {\bibnamefont {Sire}, \bibfnamefont {C.}}, \bibinfo
  {author} {\bibnamefont {Hemelrijk}, \bibfnamefont {C.~K.}}, and\ \bibinfo
  {author} {\bibnamefont {Theraulaz}, \bibfnamefont {G.}},\ }\bibfield  {title}
  {\enquote {\bibinfo {title} {{Social conformity and propagation of
  information in collective u-turns of fish schools}},}\ }\href
  {https://doi.org/10.1098/rspb.2018.0251} {\bibfield  {journal} {\bibinfo
  {journal} {Proceedings of the Royal Society B: Biological Sciences}\ }\textbf
  {\bibinfo {volume} {285}} (\bibinfo {year} {2018}),\
  10.1098/rspb.2018.0251}\BibitemShut {NoStop}%
\bibitem [{\citenamefont {Li}\ \emph {et~al.}(2021{\natexlab{a}})\citenamefont
  {Li}, \citenamefont {Liu}, \citenamefont {Deng}, \citenamefont {Lutz},\ and\
  \citenamefont {Xie}}]{Li2021a}%
  \BibitemOpen
  \bibfield  {author} {\bibinfo {author} {\bibnamefont {Li}, \bibfnamefont
  {L.}}, \bibinfo {author} {\bibnamefont {Liu}, \bibfnamefont {D.}}, \bibinfo
  {author} {\bibnamefont {Deng}, \bibfnamefont {J.}}, \bibinfo {author}
  {\bibnamefont {Lutz}, \bibfnamefont {M.~J.}}, and\ \bibinfo {author}
  {\bibnamefont {Xie}, \bibfnamefont {G.}},\ }\bibfield  {title} {\enquote
  {\bibinfo {title} {{Fish can save energy via proprioceptive sensing}},}\
  }\href {https://doi.org/10.1088/1748-3190/ac165e} {\bibfield  {journal}
  {\bibinfo  {journal} {Bioinspiration and Biomimetics}\ }\textbf {\bibinfo
  {volume} {16}} (\bibinfo {year} {2021}{\natexlab{a}}),\
  10.1088/1748-3190/ac165e}\BibitemShut {NoStop}%
\bibitem [{\citenamefont {Li}\ \emph {et~al.}(2020)\citenamefont {Li},
  \citenamefont {Nagy}, \citenamefont {Graving}, \citenamefont {Bak-Coleman},
  \citenamefont {Xie},\ and\ \citenamefont {Couzin}}]{Li2020}%
  \BibitemOpen
  \bibfield  {author} {\bibinfo {author} {\bibnamefont {Li}, \bibfnamefont
  {L.}}, \bibinfo {author} {\bibnamefont {Nagy}, \bibfnamefont {M.}}, \bibinfo
  {author} {\bibnamefont {Graving}, \bibfnamefont {J.~M.}}, \bibinfo {author}
  {\bibnamefont {Bak-Coleman}, \bibfnamefont {J.}}, \bibinfo {author}
  {\bibnamefont {Xie}, \bibfnamefont {G.}}, and\ \bibinfo {author}
  {\bibnamefont {Couzin}, \bibfnamefont {I.~D.}},\ }\bibfield  {title}
  {\enquote {\bibinfo {title} {{Vortex phase matching as a strategy for
  schooling in robots and in fish}},}\ }\href
  {https://doi.org/10.1038/s41467-020-19086-0} {\bibfield  {journal} {\bibinfo
  {journal} {Nature Communications}\ } (\bibinfo {year} {2020}),\
  10.1038/s41467-020-19086-0}\BibitemShut {NoStop}%
\bibitem [{\citenamefont {Li}\ \emph {et~al.}(2021{\natexlab{b}})\citenamefont
  {Li}, \citenamefont {Ravi}, \citenamefont {Xie},\ and\ \citenamefont
  {Couzin}}]{Li2021}%
  \BibitemOpen
  \bibfield  {author} {\bibinfo {author} {\bibnamefont {Li}, \bibfnamefont
  {L.}}, \bibinfo {author} {\bibnamefont {Ravi}, \bibfnamefont {S.}}, \bibinfo
  {author} {\bibnamefont {Xie}, \bibfnamefont {G.}}, and\ \bibinfo {author}
  {\bibnamefont {Couzin}, \bibfnamefont {I.~D.}},\ }\bibfield  {title}
  {\enquote {\bibinfo {title} {{Using a robotic platform to study the influence
  of relative tailbeat phase on the energetic costs of side-by-side swimming in
  fish}},}\ }\href {https://doi.org/10.1098/rspa.2020.0810} {\bibfield
  {journal} {\bibinfo  {journal} {Proceedings of the Royal Society A:
  Mathematical, Physical and Engineering Sciences}\ }\textbf {\bibinfo {volume}
  {477}} (\bibinfo {year} {2021}{\natexlab{b}}),\
  10.1098/rspa.2020.0810}\BibitemShut {NoStop}%
\bibitem [{\citenamefont {Lin}\ \emph {et~al.}(2023{\natexlab{a}})\citenamefont
  {Lin}, \citenamefont {Bhalla}, \citenamefont {Griffith}, \citenamefont
  {Sheng}, \citenamefont {Li}, \citenamefont {Liang},\ and\ \citenamefont
  {Zhang}}]{lin2022swimming}%
  \BibitemOpen
  \bibfield  {author} {\bibinfo {author} {\bibnamefont {Lin}, \bibfnamefont
  {Z.}}, \bibinfo {author} {\bibnamefont {Bhalla}, \bibfnamefont {A.~P.~S.}},
  \bibinfo {author} {\bibnamefont {Griffith}, \bibfnamefont {B.~E.}}, \bibinfo
  {author} {\bibnamefont {Sheng}, \bibfnamefont {Z.}}, \bibinfo {author}
  {\bibnamefont {Li}, \bibfnamefont {H.}}, \bibinfo {author} {\bibnamefont
  {Liang}, \bibfnamefont {D.}}, and\ \bibinfo {author} {\bibnamefont {Zhang},
  \bibfnamefont {Y.}},\ }\bibfield  {title} {\enquote {\bibinfo {title} {{How
  swimming style and schooling affect the hydrodynamics of two accelerating
  wavy hydrofoils}},}\ }\href
  {https://doi.org/https://doi.org/10.1016/j.oceaneng.2022.113314} {\bibfield
  {journal} {\bibinfo  {journal} {Ocean Engineering}\ }\textbf {\bibinfo
  {volume} {268}},\ \bibinfo {pages} {113314} (\bibinfo {year}
  {2023}{\natexlab{a}})}\BibitemShut {NoStop}%
\bibitem [{\citenamefont {Lin}\ \emph {et~al.}(2023{\natexlab{b}})\citenamefont
  {Lin}, \citenamefont {Liang}, \citenamefont {Bhalla}, \citenamefont
  {Al-Shabab}, \citenamefont {Skote}, \citenamefont {Zheng},\ and\
  \citenamefont {Zhang}}]{lin2023wavelength}%
  \BibitemOpen
  \bibfield  {author} {\bibinfo {author} {\bibnamefont {Lin}, \bibfnamefont
  {Z.}}, \bibinfo {author} {\bibnamefont {Liang}, \bibfnamefont {D.}}, \bibinfo
  {author} {\bibnamefont {Bhalla}, \bibfnamefont {A.~P.~S.}}, \bibinfo {author}
  {\bibnamefont {Al-Shabab}, \bibfnamefont {A.~A.~S.}}, \bibinfo {author}
  {\bibnamefont {Skote}, \bibfnamefont {M.}}, \bibinfo {author} {\bibnamefont
  {Zheng}, \bibfnamefont {W.}}, and\ \bibinfo {author} {\bibnamefont {Zhang},
  \bibfnamefont {Y.}},\ }\bibfield  {title} {\enquote {\bibinfo {title} {{How
  wavelength affects hydrodynamic performance of two accelerating
  mirror-symmetric undulating hydrofoils}},}\ }\href@noop {} {\bibfield
  {journal} {\bibinfo  {journal} {Physics of Fluids}\ } (\bibinfo {year}
  {2023}{\natexlab{b}})}\BibitemShut {NoStop}%
\bibitem [{\citenamefont {Lin}, \citenamefont {Liang},\ and\ \citenamefont
  {Zhao}(2016)}]{Lin2016a}%
  \BibitemOpen
  \bibfield  {author} {\bibinfo {author} {\bibnamefont {Lin}, \bibfnamefont
  {Z.}}, \bibinfo {author} {\bibnamefont {Liang}, \bibfnamefont {D.}}, and\
  \bibinfo {author} {\bibnamefont {Zhao}, \bibfnamefont {M.}},\ }\bibfield
  {title} {\enquote {\bibinfo {title} {{Numerical Study of the Interaction
  Between Two Immersed Cylinders}},}\ }in\ \href@noop {} {\emph {\bibinfo
  {booktitle} {The 12th International Conference on Hydrodynamics}}}\ (\bibinfo
  {year} {2016})\ p.~\bibinfo {pages} {55}\BibitemShut {NoStop}%
\bibitem [{\citenamefont {Lin}, \citenamefont {Liang},\ and\ \citenamefont
  {Zhao}(2017)}]{Lin2017a}%
  \BibitemOpen
  \bibfield  {author} {\bibinfo {author} {\bibnamefont {Lin}, \bibfnamefont
  {Z.}}, \bibinfo {author} {\bibnamefont {Liang}, \bibfnamefont {D.}}, and\
  \bibinfo {author} {\bibnamefont {Zhao}, \bibfnamefont {M.}},\ }\bibfield
  {title} {\enquote {\bibinfo {title} {{Interaction Between Two Vibrating
  Cylinders Immersed in Fluid}},}\ }in\ \href@noop {} {\emph {\bibinfo
  {booktitle} {The 27th International Ocean and Polar Engineering
  Conference}}}\ (\bibinfo  {publisher} {International Society of Offshore and
  Polar Engineers},\ \bibinfo {address} {San Francisco, California, USA},\
  \bibinfo {year} {2017})\ p.~\bibinfo {pages} {8}\BibitemShut {NoStop}%
\bibitem [{\citenamefont {Lin}, \citenamefont {Liang},\ and\ \citenamefont
  {Zhao}(2018{\natexlab{a}})}]{Lin2018b}%
  \BibitemOpen
  \bibfield  {author} {\bibinfo {author} {\bibnamefont {Lin}, \bibfnamefont
  {Z.}}, \bibinfo {author} {\bibnamefont {Liang}, \bibfnamefont {D.}}, and\
  \bibinfo {author} {\bibnamefont {Zhao}, \bibfnamefont {M.}},\ }\bibfield
  {title} {\enquote {\bibinfo {title} {{Effects of Damping on Flow-Mediated
  Interaction Between Two Cylinders}},}\ }\href
  {https://doi.org/10.1115/1.4039712} {\bibfield  {journal} {\bibinfo
  {journal} {Journal of Fluids Engineering - ASME}\ } (\bibinfo {year}
  {2018}{\natexlab{a}}),\ 10.1115/1.4039712}\BibitemShut {NoStop}%
\bibitem [{\citenamefont {Lin}, \citenamefont {Liang},\ and\ \citenamefont
  {Zhao}(2018{\natexlab{b}})}]{Lin2018c}%
  \BibitemOpen
  \bibfield  {author} {\bibinfo {author} {\bibnamefont {Lin}, \bibfnamefont
  {Z.}}, \bibinfo {author} {\bibnamefont {Liang}, \bibfnamefont {D.}}, and\
  \bibinfo {author} {\bibnamefont {Zhao}, \bibfnamefont {M.}},\ }\bibfield
  {title} {\enquote {\bibinfo {title} {{Flow-Mediated Interaction Between a
  Vibrating Cylinder and an Elastically-Mounted Cylinder}},}\ }\href
  {https://doi.org/10.1016/j.oceaneng.2018.04.019} {\bibfield  {journal}
  {\bibinfo  {journal} {Ocean Engineering}\ } (\bibinfo {year}
  {2018}{\natexlab{b}}),\ 10.1016/j.oceaneng.2018.04.019}\BibitemShut {NoStop}%
\bibitem [{\citenamefont {Lin}, \citenamefont {Liang},\ and\ \citenamefont
  {Zhao}(2019)}]{Lin2019}%
  \BibitemOpen
  \bibfield  {author} {\bibinfo {author} {\bibnamefont {Lin}, \bibfnamefont
  {Z.}}, \bibinfo {author} {\bibnamefont {Liang}, \bibfnamefont {D.}}, and\
  \bibinfo {author} {\bibnamefont {Zhao}, \bibfnamefont {M.}},\ }\bibfield
  {title} {\enquote {\bibinfo {title} {{Effects of Reynolds Number on
  Flow-Mediated Interaction between Two Cylinders}},}\ }\href
  {https://doi.org/10.1061/(ASCE)EM.1943-7889.0001670} {\bibfield  {journal}
  {\bibinfo  {journal} {Journal of Engineering Mechanics - ASCE}\ }\textbf
  {\bibinfo {volume} {145}},\ \bibinfo {pages} {04019104} (\bibinfo {year}
  {2019})}\BibitemShut {NoStop}%
\bibitem [{\citenamefont {Lin}, \citenamefont {Liang},\ and\ \citenamefont
  {Zhao}(2022)}]{lin2022flow}%
  \BibitemOpen
  \bibfield  {author} {\bibinfo {author} {\bibnamefont {Lin}, \bibfnamefont
  {Z.}}, \bibinfo {author} {\bibnamefont {Liang}, \bibfnamefont {D.}}, and\
  \bibinfo {author} {\bibnamefont {Zhao}, \bibfnamefont {M.}},\ }\bibfield
  {title} {\enquote {\bibinfo {title} {{Flow-Mediated Interaction Between a
  Forced-Oscillating Cylinder and an Elastically-Mounted Cylinder in Less
  Regular Regimes}},}\ }\href@noop {} {\bibfield  {journal} {\bibinfo
  {journal} {Physics of Fluids}\ } (\bibinfo {year} {2022})}\BibitemShut
  {NoStop}%
\bibitem [{\citenamefont {Lin}\ and\ \citenamefont
  {Zhang}(2022)}]{lin2022wavelength}%
  \BibitemOpen
  \bibfield  {author} {\bibinfo {author} {\bibnamefont {Lin}, \bibfnamefont
  {Z.}}and\ \bibinfo {author} {\bibnamefont {Zhang}, \bibfnamefont {Y.}},\
  }\bibfield  {title} {\enquote {\bibinfo {title} {{How wavelength affects the
  hydrodynamic performance of two accelerating mirror-symmetric slender
  swimmers}},}\ }\href@noop {} {\bibfield  {journal} {\bibinfo  {journal}
  {arXiv preprint arXiv:2212.11004}\ } (\bibinfo {year} {2022})}\BibitemShut
  {NoStop}%
\bibitem [{\citenamefont {Lindsey}(1978)}]{Lindsey1978}%
  \BibitemOpen
  \bibfield  {author} {\bibinfo {author} {\bibnamefont {Lindsey}, \bibfnamefont
  {C.~C.}},\ }\bibfield  {title} {\enquote {\bibinfo {title} {{Form, function,
  and locomotory habits in fish}},}\ }in\ \href
  {https://doi.org/10.1016/S1546-5098(08)60163-6} {\emph {\bibinfo {booktitle}
  {Fish Physiology}}},\ Vol.~\bibinfo {volume} {7}\ (\bibinfo {year}
  {1978})\BibitemShut {NoStop}%
\bibitem [{\citenamefont {Moriche}, \citenamefont {Flores},\ and\ \citenamefont
  {Garc{\'{i}}a-Villalba}(2016)}]{Moriche2016}%
  \BibitemOpen
  \bibfield  {author} {\bibinfo {author} {\bibnamefont {Moriche}, \bibfnamefont
  {M.}}, \bibinfo {author} {\bibnamefont {Flores}, \bibfnamefont {O.}}, and\
  \bibinfo {author} {\bibnamefont {Garc{\'{i}}a-Villalba}, \bibfnamefont
  {M.}},\ }\bibfield  {title} {\enquote {\bibinfo {title} {{Three-dimensional
  instabilities in the wake of a flapping wing at low Reynolds number}},}\
  }\href {https://doi.org/10.1016/j.ijheatfluidflow.2016.06.015} {\bibfield
  {journal} {\bibinfo  {journal} {International Journal of Heat and Fluid
  Flow}\ }\textbf {\bibinfo {volume} {62}} (\bibinfo {year} {2016}),\
  10.1016/j.ijheatfluidflow.2016.06.015}\BibitemShut {NoStop}%
\bibitem [{\citenamefont {Nair}\ and\ \citenamefont {Kanso}(2007)}]{NAIR2007}%
  \BibitemOpen
  \bibfield  {author} {\bibinfo {author} {\bibnamefont {Nair}, \bibfnamefont
  {S.}}and\ \bibinfo {author} {\bibnamefont {Kanso}, \bibfnamefont {E.}},\
  }\bibfield  {title} {\enquote {\bibinfo {title} {{Hydrodynamically Coupled
  Rigid Bodies}},}\ }\href {https://doi.org/10.1017/S002211200700849X}
  {\bibfield  {journal} {\bibinfo  {journal} {Journal of Fluid Mechanics}\
  }\textbf {\bibinfo {volume} {592}},\ \bibinfo {pages} {393--411} (\bibinfo
  {year} {2007})}\BibitemShut {NoStop}%
\bibitem [{\citenamefont {Nangia}\ \emph {et~al.}(2017)\citenamefont {Nangia},
  \citenamefont {Johansen}, \citenamefont {Patankar},\ and\ \citenamefont
  {Bhalla}}]{Nangia2017}%
  \BibitemOpen
  \bibfield  {author} {\bibinfo {author} {\bibnamefont {Nangia}, \bibfnamefont
  {N.}}, \bibinfo {author} {\bibnamefont {Johansen}, \bibfnamefont {H.}},
  \bibinfo {author} {\bibnamefont {Patankar}, \bibfnamefont {N.~A.}}, and\
  \bibinfo {author} {\bibnamefont {Bhalla}, \bibfnamefont {A.~P.~S.}},\
  }\bibfield  {title} {\enquote {\bibinfo {title} {{A moving control volume
  approach to computing hydrodynamic forces and torques on immersed bodies}},}\
  }\href {https://doi.org/10.1016/j.jcp.2017.06.047} {\bibfield  {journal}
  {\bibinfo  {journal} {Journal of Computational Physics}\ }\textbf {\bibinfo
  {volume} {347}} (\bibinfo {year} {2017}),\
  10.1016/j.jcp.2017.06.047}\BibitemShut {NoStop}%
\bibitem [{\citenamefont {Nangia}, \citenamefont {Patankar},\ and\
  \citenamefont {Bhalla}(2019)}]{Nangia2019}%
  \BibitemOpen
  \bibfield  {author} {\bibinfo {author} {\bibnamefont {Nangia}, \bibfnamefont
  {N.}}, \bibinfo {author} {\bibnamefont {Patankar}, \bibfnamefont {N.~A.}},
  and\ \bibinfo {author} {\bibnamefont {Bhalla}, \bibfnamefont {A.~P.~S.}},\
  }\bibfield  {title} {\enquote {\bibinfo {title} {{A DLM immersed boundary
  method based wave-structure interaction solver for high density ratio
  multiphase flows}},}\ }\href {https://doi.org/10.1016/j.jcp.2019.07.004}
  {\bibfield  {journal} {\bibinfo  {journal} {Journal of Computational
  Physics}\ }\textbf {\bibinfo {volume} {398}} (\bibinfo {year} {2019}),\
  10.1016/j.jcp.2019.07.004}\BibitemShut {NoStop}%
\bibitem [{\citenamefont {Pan}\ and\ \citenamefont
  {Dong}(2022)}]{pan2022effects}%
  \BibitemOpen
  \bibfield  {author} {\bibinfo {author} {\bibnamefont {Pan}, \bibfnamefont
  {Y.}}and\ \bibinfo {author} {\bibnamefont {Dong}, \bibfnamefont {H.}},\
  }\bibfield  {title} {\enquote {\bibinfo {title} {{Effects of phase difference
  on hydrodynamic interactions and wake patterns in high-density fish
  schools}},}\ }\href@noop {} {\bibfield  {journal} {\bibinfo  {journal}
  {Physics of Fluids}\ }\textbf {\bibinfo {volume} {34}},\ \bibinfo {pages}
  {111902} (\bibinfo {year} {2022})}\BibitemShut {NoStop}%
\bibitem [{\citenamefont {Partridge}(1981)}]{Partridge1981}%
  \BibitemOpen
  \bibfield  {author} {\bibinfo {author} {\bibnamefont {Partridge},
  \bibfnamefont {B.~L.}},\ }\bibfield  {title} {\enquote {\bibinfo {title}
  {{Internal dynamics and the interrelations of fish in schools}},}\ }\href
  {https://doi.org/10.1007/BF00612563} {\bibfield  {journal} {\bibinfo
  {journal} {Journal of Comparative Physiology A}\ }\textbf {\bibinfo {volume}
  {144}} (\bibinfo {year} {1981}),\ 10.1007/BF00612563}\BibitemShut {NoStop}%
\bibitem [{\citenamefont {Rosenthal}\ \emph {et~al.}(2015)\citenamefont
  {Rosenthal}, \citenamefont {Twomey}, \citenamefont {Hartnett}, \citenamefont
  {Wu},\ and\ \citenamefont {Couzin}}]{Rosenthal2015}%
  \BibitemOpen
  \bibfield  {author} {\bibinfo {author} {\bibnamefont {Rosenthal},
  \bibfnamefont {S.~B.}}, \bibinfo {author} {\bibnamefont {Twomey},
  \bibfnamefont {C.~R.}}, \bibinfo {author} {\bibnamefont {Hartnett},
  \bibfnamefont {A.~T.}}, \bibinfo {author} {\bibnamefont {Wu}, \bibfnamefont
  {H.~S.}}, and\ \bibinfo {author} {\bibnamefont {Couzin}, \bibfnamefont
  {I.~D.}},\ }\bibfield  {title} {\enquote {\bibinfo {title} {{Revealing the
  hidden networks of interaction in mobile animal groups allows prediction of
  complex behavioral contagion}},}\ }\href
  {https://doi.org/10.1073/pnas.1420068112} {\bibfield  {journal} {\bibinfo
  {journal} {Proceedings of the National Academy of Sciences of the United
  States of America}\ }\textbf {\bibinfo {volume} {112}} (\bibinfo {year}
  {2015}),\ 10.1073/pnas.1420068112}\BibitemShut {NoStop}%
\bibitem [{\citenamefont {Santo}\ \emph {et~al.}(2021)\citenamefont {Santo},
  \citenamefont {Goerig}, \citenamefont {Wainwright}, \citenamefont {Akanyeti},
  \citenamefont {Liao}, \citenamefont {Castro-Santos},\ and\ \citenamefont
  {Lauder}}]{Santo2021}%
  \BibitemOpen
  \bibfield  {author} {\bibinfo {author} {\bibnamefont {Santo}, \bibfnamefont
  {V.~D.}}, \bibinfo {author} {\bibnamefont {Goerig}, \bibfnamefont {E.}},
  \bibinfo {author} {\bibnamefont {Wainwright}, \bibfnamefont {D.~K.}},
  \bibinfo {author} {\bibnamefont {Akanyeti}, \bibfnamefont {O.}}, \bibinfo
  {author} {\bibnamefont {Liao}, \bibfnamefont {J.~C.}}, \bibinfo {author}
  {\bibnamefont {Castro-Santos}, \bibfnamefont {T.}}, and\ \bibinfo {author}
  {\bibnamefont {Lauder}, \bibfnamefont {G.~V.}},\ }\bibfield  {title}
  {\enquote {\bibinfo {title} {{Convergence of undulatory swimming kinematics
  across a diversity of fishes}},}\ }\href
  {https://doi.org/10.1073/pnas.2113206118} {\bibfield  {journal} {\bibinfo
  {journal} {Proceedings of the National Academy of Sciences of the United
  States of America}\ }\textbf {\bibinfo {volume} {118}} (\bibinfo {year}
  {2021}),\ 10.1073/pnas.2113206118}\BibitemShut {NoStop}%
\bibitem [{\citenamefont {Schwalbe}\ \emph {et~al.}(2019)\citenamefont
  {Schwalbe}, \citenamefont {Boden}, \citenamefont {Wise},\ and\ \citenamefont
  {Tytell}}]{Schwalbe2019}%
  \BibitemOpen
  \bibfield  {author} {\bibinfo {author} {\bibnamefont {Schwalbe},
  \bibfnamefont {M.~A.}}, \bibinfo {author} {\bibnamefont {Boden},
  \bibfnamefont {A.~L.}}, \bibinfo {author} {\bibnamefont {Wise}, \bibfnamefont
  {T.~N.}}, and\ \bibinfo {author} {\bibnamefont {Tytell}, \bibfnamefont
  {E.~D.}},\ }\bibfield  {title} {\enquote {\bibinfo {title} {{Red muscle
  activity in bluegill sunfish Lepomis macrochirus during forward
  accelerations}},}\ }\href {https://doi.org/10.1038/s41598-019-44409-7}
  {\bibfield  {journal} {\bibinfo  {journal} {Scientific Reports}\ }\textbf
  {\bibinfo {volume} {9}} (\bibinfo {year} {2019}),\
  10.1038/s41598-019-44409-7}\BibitemShut {NoStop}%
\bibitem [{\citenamefont {Sfakiotakis}, \citenamefont {Lane},\ and\
  \citenamefont {Davies}(1999)}]{Sfakiotakis1999}%
  \BibitemOpen
  \bibfield  {author} {\bibinfo {author} {\bibnamefont {Sfakiotakis},
  \bibfnamefont {M.}}, \bibinfo {author} {\bibnamefont {Lane}, \bibfnamefont
  {D.~M.}}, and\ \bibinfo {author} {\bibnamefont {Davies}, \bibfnamefont
  {J.~B.~C.}},\ }\bibfield  {title} {\enquote {\bibinfo {title} {{Review of
  fish swimming modes for aquatic locomotion}},}\ }\href
  {https://doi.org/10.1109/48.757275} {\bibfield  {journal} {\bibinfo
  {journal} {IEEE Journal of Oceanic Engineering}\ }\textbf {\bibinfo {volume}
  {24}} (\bibinfo {year} {1999}),\ 10.1109/48.757275}\BibitemShut {NoStop}%
\bibitem [{\citenamefont {Shao}\ \emph {et~al.}(2010)\citenamefont {Shao},
  \citenamefont {Pan}, \citenamefont {Deng},\ and\ \citenamefont
  {Yu}}]{Shao2010}%
  \BibitemOpen
  \bibfield  {author} {\bibinfo {author} {\bibnamefont {Shao}, \bibfnamefont
  {X.}}, \bibinfo {author} {\bibnamefont {Pan}, \bibfnamefont {D.}}, \bibinfo
  {author} {\bibnamefont {Deng}, \bibfnamefont {J.}}, and\ \bibinfo {author}
  {\bibnamefont {Yu}, \bibfnamefont {Z.}},\ }\bibfield  {title} {\enquote
  {\bibinfo {title} {{Hydrodynamic performance of a fishlike undulating foil in
  the wake of a cylinder}},}\ }\href {https://doi.org/10.1063/1.3504651}
  {\bibfield  {journal} {\bibinfo  {journal} {Physics of Fluids}\ }\textbf
  {\bibinfo {volume} {22}} (\bibinfo {year} {2010}),\
  10.1063/1.3504651}\BibitemShut {NoStop}%
\bibitem [{\citenamefont {Shrivastava}\ \emph {et~al.}(2017)\citenamefont
  {Shrivastava}, \citenamefont {Malushte}, \citenamefont {Agrawal},\ and\
  \citenamefont {Sharma}}]{Shrivastava2017}%
  \BibitemOpen
  \bibfield  {author} {\bibinfo {author} {\bibnamefont {Shrivastava},
  \bibfnamefont {M.}}, \bibinfo {author} {\bibnamefont {Malushte},
  \bibfnamefont {M.}}, \bibinfo {author} {\bibnamefont {Agrawal}, \bibfnamefont
  {A.}}, and\ \bibinfo {author} {\bibnamefont {Sharma}, \bibfnamefont {A.}},\
  }\bibfield  {title} {\enquote {\bibinfo {title} {{CFD study on hydrodynamics
  of three fish-like undulating hydrofoils in side-by-side arrangement}},}\
  }\href@noop {} {\bibfield  {journal} {\bibinfo  {journal} {Fluid Mechanics
  and Fluid Power – Contemporary Research}\ } (\bibinfo {year}
  {2017})}\BibitemShut {NoStop}%
\bibitem [{\citenamefont {Thekkethil}, \citenamefont {Sharma},\ and\
  \citenamefont {Agrawal}(2018)}]{Thekkethil2018}%
  \BibitemOpen
  \bibfield  {author} {\bibinfo {author} {\bibnamefont {Thekkethil},
  \bibfnamefont {N.}}, \bibinfo {author} {\bibnamefont {Sharma}, \bibfnamefont
  {A.}}, and\ \bibinfo {author} {\bibnamefont {Agrawal}, \bibfnamefont {A.}},\
  }\bibfield  {title} {\enquote {\bibinfo {title} {{Unified hydrodynamics study
  for various types of fishes-like undulating rigid hydrofoil in a free stream
  flow}},}\ }\href {https://doi.org/10.1063/1.5041358} {\bibfield  {journal}
  {\bibinfo  {journal} {Physics of Fluids}\ }\textbf {\bibinfo {volume} {30}}
  (\bibinfo {year} {2018}),\ 10.1063/1.5041358}\BibitemShut {NoStop}%
\bibitem [{\citenamefont {Thekkethil}, \citenamefont {Sharma},\ and\
  \citenamefont {Agrawal}(2020)}]{Thekkethil2020}%
  \BibitemOpen
  \bibfield  {author} {\bibinfo {author} {\bibnamefont {Thekkethil},
  \bibfnamefont {N.}}, \bibinfo {author} {\bibnamefont {Sharma}, \bibfnamefont
  {A.}}, and\ \bibinfo {author} {\bibnamefont {Agrawal}, \bibfnamefont {A.}},\
  }\bibfield  {title} {\enquote {\bibinfo {title} {{Self-propulsion of
  fishes-like undulating hydrofoil: A unified kinematics based unsteady
  hydrodynamics study}},}\ }\href
  {https://doi.org/10.1016/j.jfluidstructs.2020.102875} {\bibfield  {journal}
  {\bibinfo  {journal} {Journal of Fluids and Structures}\ }\textbf {\bibinfo
  {volume} {93}} (\bibinfo {year} {2020}),\
  10.1016/j.jfluidstructs.2020.102875}\BibitemShut {NoStop}%
\bibitem [{\citenamefont {Thekkethil}\ \emph {et~al.}(2017)\citenamefont
  {Thekkethil}, \citenamefont {Shrivastava}, \citenamefont {Agrawal},\ and\
  \citenamefont {Sharma}}]{Thekkethil2017}%
  \BibitemOpen
  \bibfield  {author} {\bibinfo {author} {\bibnamefont {Thekkethil},
  \bibfnamefont {N.}}, \bibinfo {author} {\bibnamefont {Shrivastava},
  \bibfnamefont {M.}}, \bibinfo {author} {\bibnamefont {Agrawal}, \bibfnamefont
  {A.}}, and\ \bibinfo {author} {\bibnamefont {Sharma}, \bibfnamefont {A.}},\
  }\bibfield  {title} {\enquote {\bibinfo {title} {{Effect of wavelength of
  fish-like undulation of a hydrofoil in a free-stream flow}},}\ }\href
  {https://doi.org/10.1007/s12046-017-0619-7} {\bibfield  {journal} {\bibinfo
  {journal} {Sadhana - Academy Proceedings in Engineering Sciences}\ }\textbf
  {\bibinfo {volume} {42}} (\bibinfo {year} {2017}),\
  10.1007/s12046-017-0619-7}\BibitemShut {NoStop}%
\bibitem [{\citenamefont {Tytell}(2004)}]{Tytell2004b}%
  \BibitemOpen
  \bibfield  {author} {\bibinfo {author} {\bibnamefont {Tytell}, \bibfnamefont
  {E.~D.}},\ }\bibfield  {title} {\enquote {\bibinfo {title} {{Kinematics and
  hydrodynamics of linear acceleration in eels, Anguilla rostrata}},}\ }\href
  {https://doi.org/10.1098/rspb.2004.2901} {\bibfield  {journal} {\bibinfo
  {journal} {Proceedings of the Royal Society B: Biological Sciences}\ }\textbf
  {\bibinfo {volume} {271}} (\bibinfo {year} {2004}),\
  10.1098/rspb.2004.2901}\BibitemShut {NoStop}%
\bibitem [{\citenamefont {Webb}(1984)}]{Webb1984}%
  \BibitemOpen
  \bibfield  {author} {\bibinfo {author} {\bibnamefont {Webb}, \bibfnamefont
  {P.~W.}},\ }\bibfield  {title} {\enquote {\bibinfo {title} {{Form and
  Function in Fish Swimming}},}\ }\href
  {https://doi.org/10.1038/scientificamerican0784-72} {\bibfield  {journal}
  {\bibinfo  {journal} {Scientific American}\ }\textbf {\bibinfo {volume}
  {251}} (\bibinfo {year} {1984}),\
  10.1038/scientificamerican0784-72}\BibitemShut {NoStop}%
\bibitem [{\citenamefont {Wei}\ \emph {et~al.}(2022)\citenamefont {Wei},
  \citenamefont {Hu}, \citenamefont {Zhang},\ and\ \citenamefont
  {Zeng}}]{Wei2022}%
  \BibitemOpen
  \bibfield  {author} {\bibinfo {author} {\bibnamefont {Wei}, \bibfnamefont
  {C.}}, \bibinfo {author} {\bibnamefont {Hu}, \bibfnamefont {Q.}}, \bibinfo
  {author} {\bibnamefont {Zhang}, \bibfnamefont {T.}}, and\ \bibinfo {author}
  {\bibnamefont {Zeng}, \bibfnamefont {Y.}},\ }\bibfield  {title} {\enquote
  {\bibinfo {title} {{Passive hydrodynamic interactions in minimal fish
  schools}},}\ }\href {https://doi.org/10.1016/j.oceaneng.2022.110574}
  {\bibfield  {journal} {\bibinfo  {journal} {Ocean Engineering}\ }\textbf
  {\bibinfo {volume} {247}} (\bibinfo {year} {2022}),\
  10.1016/j.oceaneng.2022.110574}\BibitemShut {NoStop}%
\bibitem [{\citenamefont {Weihs}(1973)}]{weihs1973hydromechanics}%
  \BibitemOpen
  \bibfield  {author} {\bibinfo {author} {\bibnamefont {Weihs}, \bibfnamefont
  {D.}},\ }\bibfield  {title} {\enquote {\bibinfo {title} {{Hydromechanics of
  Fish Schooling}},}\ }\href@noop {} {\bibfield  {journal} {\bibinfo  {journal}
  {Nature}\ }\textbf {\bibinfo {volume} {241}},\ \bibinfo {pages} {290--291}
  (\bibinfo {year} {1973})}\BibitemShut {NoStop}%
\bibitem [{\citenamefont {Yu}\ and\ \citenamefont
  {Huang}(2021)}]{yu2021scaling}%
  \BibitemOpen
  \bibfield  {author} {\bibinfo {author} {\bibnamefont {Yu}, \bibfnamefont
  {Y.-L.}}and\ \bibinfo {author} {\bibnamefont {Huang}, \bibfnamefont
  {K.-J.}},\ }\bibfield  {title} {\enquote {\bibinfo {title} {{Scaling law of
  fish undulatory propulsion}},}\ }\href@noop {} {\bibfield  {journal}
  {\bibinfo  {journal} {Physics of Fluids}\ }\textbf {\bibinfo {volume} {33}},\
  \bibinfo {pages} {61905} (\bibinfo {year} {2021})}\BibitemShut {NoStop}%
\bibitem [{\citenamefont {Yucel}, \citenamefont {Sahin},\ and\ \citenamefont
  {Unal}(2022)}]{Yucel2022}%
  \BibitemOpen
  \bibfield  {author} {\bibinfo {author} {\bibnamefont {Yucel}, \bibfnamefont
  {S.~B.}}, \bibinfo {author} {\bibnamefont {Sahin}, \bibfnamefont {M.}}, and\
  \bibinfo {author} {\bibnamefont {Unal}, \bibfnamefont {M.~F.}},\ }\bibfield
  {title} {\enquote {\bibinfo {title} {{Propulsive performance of plunging
  airfoils in biplane configuration}},}\ }\href
  {https://doi.org/10.1063/5.0083040} {\bibfield  {journal} {\bibinfo
  {journal} {Physics of Fluids}\ }\textbf {\bibinfo {volume} {34}} (\bibinfo
  {year} {2022}),\ 10.1063/5.0083040}\BibitemShut {NoStop}%
\bibitem [{\citenamefont {Zheng}\ \emph {et~al.}(2005)\citenamefont {Zheng},
  \citenamefont {Kashimori}, \citenamefont {Hoshino}, \citenamefont {Fujita},\
  and\ \citenamefont {Kambara}}]{Zheng2005}%
  \BibitemOpen
  \bibfield  {author} {\bibinfo {author} {\bibnamefont {Zheng}, \bibfnamefont
  {M.}}, \bibinfo {author} {\bibnamefont {Kashimori}, \bibfnamefont {Y.}},
  \bibinfo {author} {\bibnamefont {Hoshino}, \bibfnamefont {O.}}, \bibinfo
  {author} {\bibnamefont {Fujita}, \bibfnamefont {K.}}, and\ \bibinfo {author}
  {\bibnamefont {Kambara}, \bibfnamefont {T.}},\ }\bibfield  {title} {\enquote
  {\bibinfo {title} {{Behavior pattern (innate action) of individuals in fish
  schools generating efficient collective evasion from predation}},}\ }\href
  {https://doi.org/10.1016/j.jtbi.2004.12.025} {\bibfield  {journal} {\bibinfo
  {journal} {Journal of Theoretical Biology}\ }\textbf {\bibinfo {volume}
  {235}} (\bibinfo {year} {2005}),\ 10.1016/j.jtbi.2004.12.025}\BibitemShut
  {NoStop}%
\end{thebibliography}%
	
	\pagebreak

	\appendix
	
	\section{Flow structure maps in detail}
	\label{appendix_flow_map_detail}

	\begin{sidewaysfigure}
		\centering
		\includegraphics[width=1\linewidth]{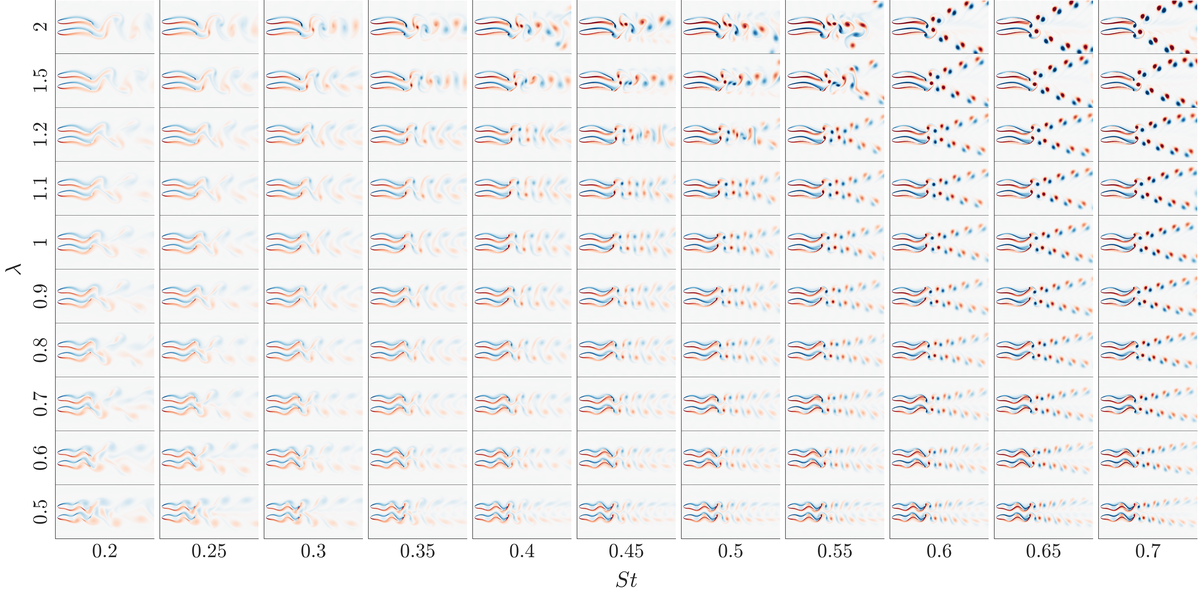}
		\caption{Flow structure visualised by vorticity contours at $ Re = 1000 $ with $ St = 0.2 - 0.7 $ and $ \lam = 0.5 - 2 $.}
		\label{fig:vor_map_re_1000}
	\end{sidewaysfigure}
	
	\begin{sidewaysfigure}
		\centering
		\includegraphics[width=1\linewidth]{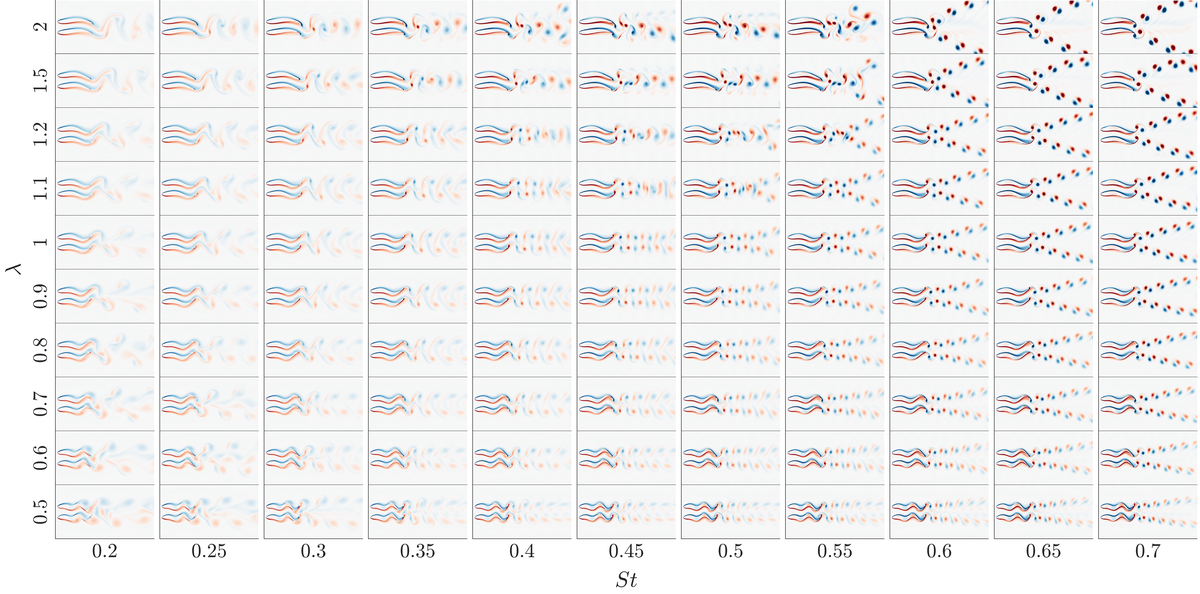}
		\caption{Flow structure visualised by vorticity contours at $ Re = 1250 $ with $ St = 0.2 - 0.7 $ and $ \lam = 0.5 - 2 $.}
		\label{fig:vor_map_re_1250}
	\end{sidewaysfigure}
	
	\begin{sidewaysfigure}
		\centering
		\includegraphics[width=1\linewidth]{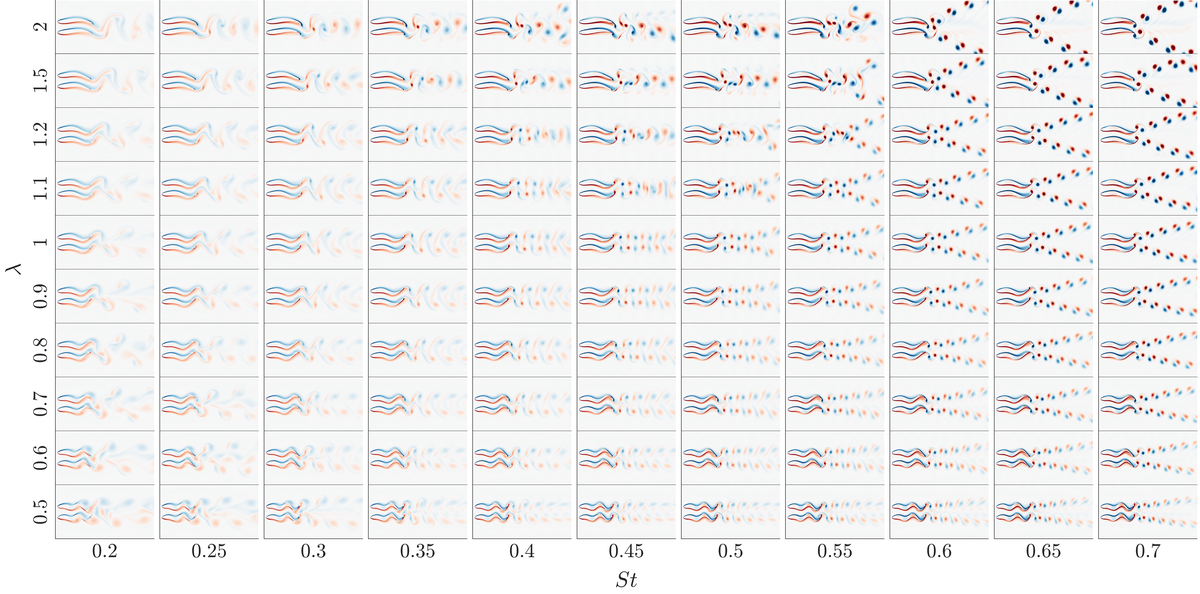}
		\caption{Flow structure visualised by vorticity contours at $ Re = 1500 $ with $ St = 0.2 - 0.7 $ and $ \lam = 0.5 - 2 $.}
		\label{fig:vor_map_re_1500}
	\end{sidewaysfigure}
	
	\begin{sidewaysfigure}
		\centering
		\includegraphics[width=1\linewidth]{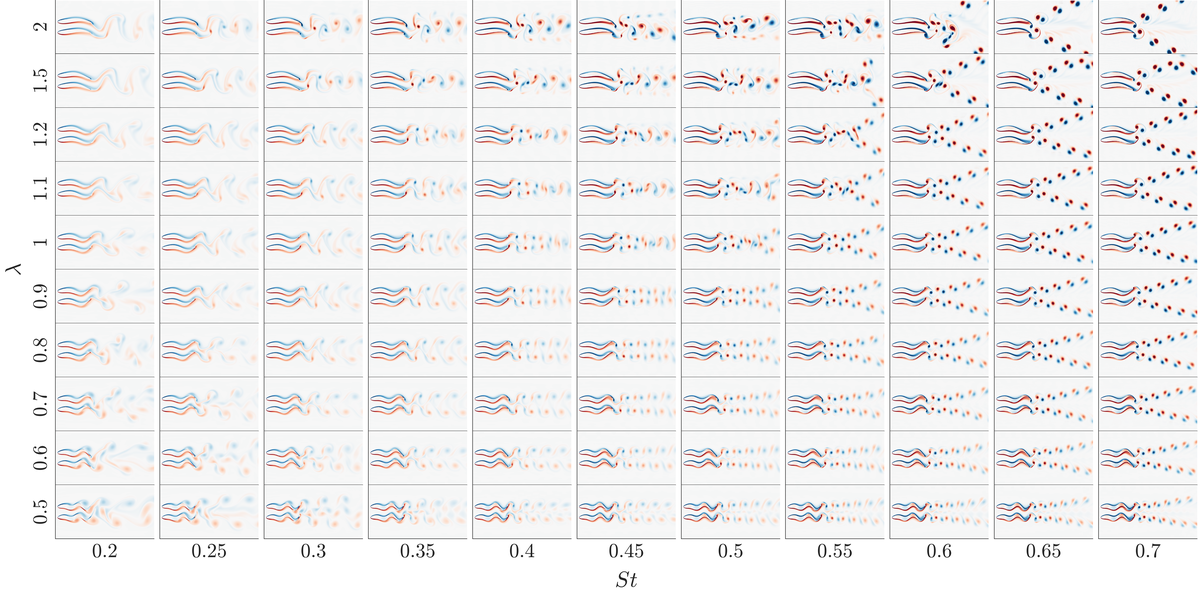}
		\caption{Flow structure visualised by vorticity contours at $ Re = 1750 $ with $ St = 0.2 - 0.7 $ and $ \lam = 0.5 - 2 $.}
		\label{fig:vor_map_re_1750}
	\end{sidewaysfigure}
	
	\begin{sidewaysfigure}
		\centering
		\includegraphics[width=1\linewidth]{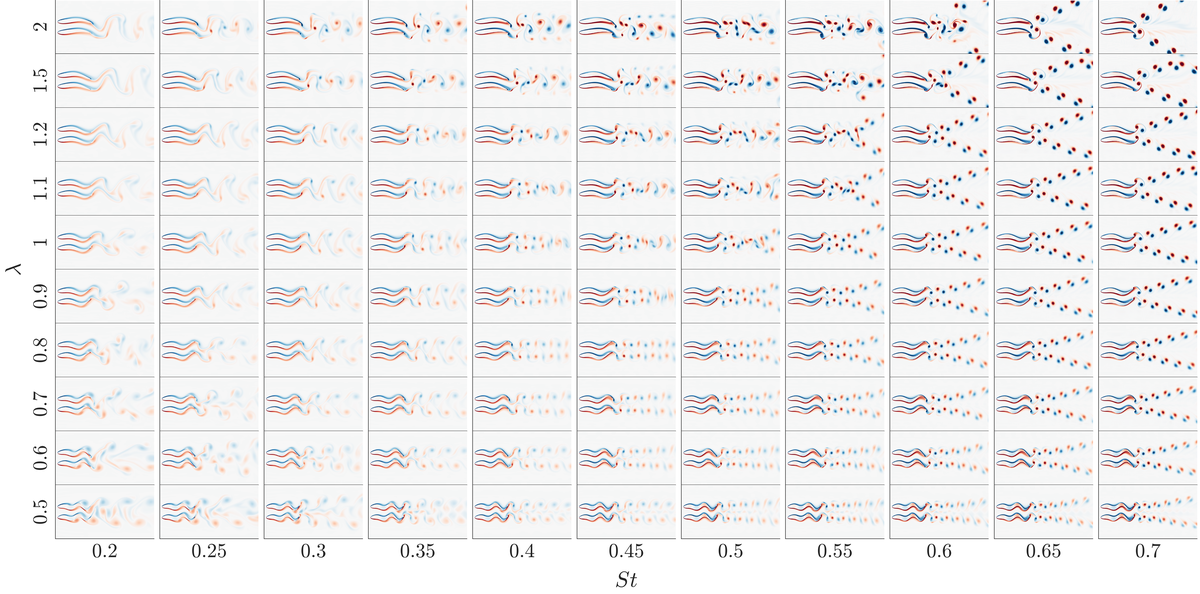}
		\caption{Flow structure visualised by vorticity contours at $ Re = 2000 $ with $ St = 0.2 - 0.7 $ and $ \lam = 0.5 - 2 $.}
		\label{fig:vor_map_re_2000}
	\end{sidewaysfigure}

\end{document}